\documentstyle[12pt]{ioplppt}

\begin{document}

\sloppy
\jl{2}
\eqnobysec


\title{
Post-Collision Interaction with  Wannier 
electrons}[PCI with  Wannier electrons]
\author{M. Yu. Kuchiev
\ftnote{2}{E-mail: kuchiev@newt.phys.unsw.edu.au} }
\address{School of Physics, The University of New South Wales,
Sydney 2052, Australia}


\begin{abstract}
A theory of the  Post-Collision Interaction (PCI)
is developed for the case
when an electron atom impact  results in  creation of two low-energy 
Wannier electrons and  an ion
excited into an autoionizing state. 
The following
autoionization decay  exposes the Wannier pair to the influence 
of PCI resulting in  variation of
the shape of the line in the autoionization spectrum. An explicit 
dependence of the autoionization profile 
on the  wave function of the Wannier pair
is found. PCI provides an opportunity  to study this wave function
for a wide area of distances.
\end{abstract}

\maketitle

\section{Introduction}
\label{intro}
This paper  develops a 
theoretical approach allowing one to consider PCI in the 
Wannier regime.
The interest in  this problem has been initiated by 
recent experiment
\cite{mel} followed by theoretical work \cite{ost} in which
the electron impact in the vicinity of the atomic inner shell
ionization threshold
\begin{equation} \label{1}
{\rm e}+ {\rm A} \rightarrow {\rm e_\alpha}+ {\rm e}_\beta + {\rm A}^{+*}
\end{equation}
 was studied. 
In this  reaction 
the inelastically scattered electron e$_\alpha$
and the knocked out electron e$_\beta$ have low energy 
propagating in the Coulomb field of the 
positive ion A$^{+*}$. 
These low-energy electrons are known as the  Wannier pair.
The Coulomb interaction of  Wannier electrons between themselves
as well as with the field of the ion 
results in  the   Wannier power
law \cite{wan} for the cross-section 
\begin{equation} \label{w1}
\sigma \sim \varepsilon^{\mu}
\end{equation}
where $\varepsilon$ 
is  the energy excess over the ionization threshold and the exponent
$\mu$ depends non-trivially on the ion charge $Z$
\begin{equation} \label{ex}
\mu =\frac{1}{4}\left[\left(\frac{100 Z - 9}{4 Z-1}\right)^{1/2} - 
1\right].
\end{equation}
A distinctive property of reaction  (\ref{1})
is the finite life-time of
the inner-shell vacancy in the  ion A$^{+*}$. Its decay
\begin{equation} \label{2}
{\rm e_\alpha + e_\beta} + 
{\rm A}^{+*} \rightarrow {\rm e_\alpha + e_\beta} + {\rm e}_\gamma 
+ {\rm A}^{2+}
\end{equation}
creates 
the fast Auger electron e$_\gamma$ and doubly-charged ion A$^{2+}$. 
The fact that the pair e$_\alpha$, e$_\beta$ has low energy guarantees 
that 
at the moment of the decay it is situated
in some close vicinity of the ion.
Therefore the pair should strongly interact
with the products of the Auger decay,
{\it i.e.} the doubly charged ion and Auger electron.
An interaction of this type is known in literature as 
the  Post-Collision Interaction (PCI). What distinguishes 
reaction (\ref{1}),(\ref{2}) 
is the  fact that the Wannier pair  takes part in  PCI.
Previously PCI was extensively studied in the near threshold
region for the case when a
single low-energy electron is localized in the vicinity
of the atom at the moment of the autoionization decay. 

Thus reaction  (\ref{1}),(\ref{2}) combines 
the two  problems of the final state interaction, 
the Wannier problem and PCI.
Previously they both had been extensively studied 
experimentally as well as theoretically
but in those situations when only one of them,
either the Wannier problem or PCI
had manifested itself in the final state. 

The main purpose of this paper is development of the
theory  describing PCI
in reaction   (\ref{1}),(\ref{2})  as well as in similar
situations in electron or atomic collisions or in photoionization. 
One of the interesting qualitative questions arising
from the reaction (\ref{1}),(\ref{2}) 
is a possible influence of PCI on 
the Wannier power law  (\ref{w1}). In the field of the singly
charged ion the Wannier exponent  (\ref{ex}) is $\mu_1 \simeq 1.127$,
while in the field
of the doubly charged ion it is $\mu_2 \simeq 1.056$. 
If the energy of the Auger electron e$_\gamma$ is high enough then 
it does not  interact strongly with the pair e$_\alpha$,e$_\beta$ and 
the decay could be considered as a switch of  the ion charge
from the intermediate-state
value $Z=1$ to the  final-state value $Z=2$.
The question is how this charge  variation affects the exponent.
We will demonstrate that 
for sufficiently low above-threshold
energy the exponent is equal to $\mu_2$, while for higher
energies it is $ \mu_1$. This fact agrees with the following simple
 physical 
picture. When above-threshold energy 
is very low then the Wannier pair is located
so close to the ion  
at the moment of the decay
that its separation 
is negligible compared
with an important in the problem radius of the Coulomb zone.
Therefore one can neglect this separation, 
neglecting  as well the very fact of existence of the 
intermediate-state resonance. 
Thus in this case the pair is influenced by the field of the 
final-state ion with the charge $Z=2$.  
In contrast for high
above-threshold energy the Wannier pair is located well outside the
radius of the Coulomb zone at the moment of the Auger decay and
therefore is not strongly influenced by the field of the 
doubly-charged ion.
The main characteristics  of the pair in this
case are governed by the ion in the intermediate state with
$Z=1$.

The paper is 
organized as follows. Section \ref{wan} 
gives a short introduction to  the Wannier problem which 
was theoretically studied in a number of papers.
The variety of  approaches developed and the comprehensive list
of references can be found in \cite{pet} - \cite{ost2}.
Section \ref{near} presents a short review of  essential ideas of PCI. 
The literature on the theory of PCI can be found  in  \cite{ks}.
These introductory notes 
permit  presenting   basic ideas of the two problems 
from a particular point of view which provides the easiest way
for further  development  in 
Sections \ref{pci} and \ref{conseq} 
where a way is  found to describe PCI 
when Wannier electrons are involved.

Generally speaking the problem considered looks very complicated
being dependent on the Coulomb interaction 
of three electrons and  the ion.
There are, however, important simplifications. Firstly, we 
assume that the Wannier pair possesses low energy, while the Auger
electron is so fast that one can neglect its interaction with
the pair. This allows us to reformulate the problem 
in terms of  the Wannier pair only.
Still it is a dynamical three-body problem.
Another important simplification
comes from the cornerstone of the Wannier problem   \cite{wan}. 
In order  to escape from the ionic
field  Wannier electrons should move along a particular
classical trajectory.
This motion can be considered as a collective mode 
describing essential properties of the system.
One can easily verify, see Section \ref{wan},
that this mode  can be described in terms of an
effective single-particle Coulomb problem. 
Of course there are important corrections arising from the three-body
nature of the problem and making it so
specific and interesting. The point is that
knowing the reliable initial approximation
one can find these corrections in a clear analytical form.

These arguments show that one should expect
the problem considered to have a clear solution.
Finding  and describing  it in simple physical terms
is the major goal of this paper.
The main result obtained  is the formula which presents the
autoionization profile    $R_{\rm pci}$  distorted by PCI in the form
\begin{equation}\label{rkr}
R_{\rm pci} = K_{\rm w} R_{\rm c}
\end{equation} 
The factor $R_{\rm c}$ here arises from the most simple
approximation for the Wannier problem described
by the above mentioned effective 
single-particle Coulomb problem. Due to this reason
we will call it the Coulomb factor. This factor depends on the 
overlapping integral of the Coulomb wave functions.
It is interesting that a similar factor 
describes  PCI in the traditional situation when there
is only one low-energy electron.
The only distinction is that in the later case 
there is a true single-particle Coulomb problem, while
for the Wannier pair 
the Coulomb problem arises as an effective approximation.
Still the similarity is so close that one can consider
the Coulomb factor $R_{\rm c}$ 
as a well known quantity which was well studied 
previously both experimentally and theoretically. 
In contrast  the factor $ K_{\rm w}$
takes into account  more subtle and interesting
properties of the process which are
essential for PCI with the Wannier pair and have no analogy
in PCI with one low-energy electron. 
Due to  this reason we will call
it the Wannier factor. In Section \ref{conseq} an explicit
analytical formula for this factor is derived. 
The Wannier factor found depends on the normalization
coefficient of the Wannier electrons wave function. It is interesting
that the Wannier power law (\ref{w1}) arises from the properties
of the same coefficient considered in the region of small
separations of the pair from the atom. 
In contrast, when PCI  is studied  this coefficient 
manifests  itself in a wide range of distances. This property of
the profile (\ref{rkr}) opens a possibility for experimental
studies of the normalization coefficient of the Wannier
pair wave function in a  wide  area of distances.

The results obtained in Sections \ref{pci},\ref{conseq} 
are illustrated in 
Section \ref{examp} for the electron impact near-threshold
 K-shell excitation of Ne. 
Section \ref{discussion} presents a short description of the major 
results.

\section{Wannier problem} 
\label{wan}
Consider two low-energy
electrons in the field of an ion with the charge $Z$.
The Coulomb repulsion between the  electrons makes  opposite
directions the most favorable for them \cite{wan}.
To simplify the following  consideration let us 
use this fact developing the following  model.
Let us assume that the electrons are moving
in opposite directions neglecting fluctuations in their
angular distribution.  
Moreover, let us  neglect as well the orbital moment.
Thus in this model 
angular variables are neglected which makes it 
essentially one-dimensional. This  model was
applied earlier for description of the Wannier power law
\cite{petlie}.
It is important that  the most interesting part of the 
Wannier problem, its instability against  a non-symmetrical
distribution of energy between the two electrons, is taken into account
in full.
Moreover, we will argue in Section \ref{discussion} that 
our final results  do not depend on  model assumptions.
One of the main advantages of the model is simplification
of notation, which inevitably is to be sufficiently complicated.

Our strategy in this Section is to first consider the classical
approximation and then 
develop the quantum picture
using the semiclassical approach which gives a correct
description of the process for low above-threshold energy.
The semiclassical approach to the Wannier problem was
developed in detail  in \cite{pet}  to describe the 
near-threshold behaviour of the reaction (\ref{1}).
In this case one is mainly interested in the
behaviour of the wave function of the Wannier pair for
small separations. We will see that to describe  PCI  it is 
necessary to know the wave function for any separation,
not necessarily in the vicinity of the atom. 
The convenient representation for this
wave function valid for an arbitrary separation is given below
in equation (\ref{answ}) which is the main result of this Section.

\subsection{Classical trajectories}
Consider classically the problem of 
two electrons e$_\alpha$,e$_\beta$ in the framework of the 
above-formulated one-dimensional model
in which electrons are supposed to move in opposite directions
with zero orbital moments. Such a system  is characterized by 
 two radii $r_\alpha,r_\beta \ge 0$. The Lagrangian in these 
coordinates is
\begin{equation} \label{lag}
L(r_\alpha,r_\beta,\dot r_\alpha,\dot r_\beta)=
\frac{\dot r_\alpha^2}{2}+\frac{\dot r_\beta^2}{2}+\frac{Z}{r_\alpha} +
\frac{Z}{r_\beta}
-\frac{1}{r_\alpha+r_\beta},
\end{equation}
where $Z$ is the charge of the ion. It is convenient to
introduce hyper-spherical  coordinates 
\begin{equation} \label{r1}
\left\{ \begin{array}{l}
r_\alpha = r \cos (\chi+\frac{\pi}{4}) \\
r_\beta = r \cos (\chi-\frac{\pi}{4})
\end{array}  \right .
\end{equation}
in which $r\ge 0$ describes the overall distance of the pair from the 
ion, and $\chi,~
- \pi/4\le \chi \le \pi/4$ shows how strongly the pair
is deviated from the symmetrical configuration $r_\alpha=r_\beta$, in 
which
$\chi=0$.
In these coordinates  the Lagrangian is
\begin{equation} \label{Lrchi} 
L(r,\chi ,\dot r,\dot \chi)=
\frac{\dot r^2}{2}+\frac{r^2 \dot \chi^2}{2} - U(r,\chi).
\end{equation}
The potential energy
\begin{equation} \label{U}
U(r,\chi) = - \frac{w(\chi)}{r}
\end{equation}
can be looked  at as a Coulomb field produced  by a  $\chi$-dependent
charge
\begin{equation} \label{W}
w(\chi) = 
\frac{(2 Z-1) \cos2\chi+2 Z}{\sqrt 2 \cos \chi \cos 2 \chi}\ge 0.
\end{equation}
The function $w(\chi)$ has a minimum for the symmetrical configuration
$\chi =0$.
It  manifests itself 
as a maximum at $\chi =0$ of the potential $U(r,\chi)$
considered as a function of $\chi$.  
This property of the potential is  known as the Wannier  ridge. 
The ridge makes the classical system
unstable. The instability results in the most important
property of the Wannier problem. 
In order to escape
 from the Coulomb field of the  
ion the two electrons should propagate 
in the vicinity of the symmetrical configuration.
Otherwise one of them would be trapped
by the ion field, {\it i.e.} the pair fall off the ridge. 
That is why the motion in the vicinity
of the top of the ridge, where the pair is stable
plays such an important role
and is considered in detail below.

In the vicinity 
of $\chi=0$ the function $w(\chi)$  can be expanded in  the Taylor series
\begin{equation} \label{w}
w(\chi) \approx w + \frac{w''}{2} \chi^2,
\end{equation}
where the coefficients are
\begin{eqnarray}\label{W02} 
&&w=\frac{4 Z-1}{\sqrt 2},\\ \label{w''}
&&w''=\frac{12 Z-1}{\sqrt 2}.
\end{eqnarray}
Equation (\ref{Lrchi}) results in equations of motion 
\begin{eqnarray}\label{newtr}
\ddot r = r\dot\chi^2 -\frac{w(\chi)}{r^2}, \\ \label{nchi}
\frac{d}{dt} (r^2 \dot \chi)= \frac{1}{r}\frac{d w}{d \chi}(\chi).
\end{eqnarray}
The   trajectory  along the top of the Wannier 
ridge  satisfies
\begin{eqnarray} \label{chi0}
\chi_0&=&0,
\\ \label{r0}
\ddot r_0 &=& - \frac{w}{r_0^2}.
\end{eqnarray}
The second equation coincides with the  equation describing the motion
of a single particle in the field of the Coulomb charge $w$. 
We will refer to this motion, which plays an important role
below,  as an effective single-particle
Coulomb problem.
The first integral of  (\ref{r0}) 
\begin{equation} \label{en0}
\frac{\dot r^2_0}{2} - \frac{w}{r_0} = \epsilon
\end{equation}
describes the energy conservation law. Using it one finds
the ridge-top trajectory
\begin{eqnarray}\label{t(r)}
t  &=& \int \frac{dr_0}{ \sqrt{ 2(\epsilon +w/r_0}) }
= 
 r_0(t)\left[2 \left( \epsilon +\frac{w}{r_0(t)}\right)\right]^{1/2}
\\ \nonumber
&+&\frac{w}{\sqrt{2 \epsilon}}
\ln \left\{ 1+ \frac{2\epsilon r_0(t)}{w}\left[ 1+ \left(1+
\frac{w}{\epsilon r_0(t)}\right)^{1/2}\right]\right\} .
\end{eqnarray}
For small deviations  from the ridge-top trajectory  
one obtains the linear equation for
$ \chi = \delta \chi$ using expansion  (\ref{w}) 
\begin{equation} \label{newtc0}
r \frac{d}{dt}(r^2 \dot \chi) =w'' \chi.
\end{equation}
To simplify notation  
the ridge-top trajectory  is denoted here by $r = r_0(t)$
defined in  (\ref{t(r)}).
For further applications we need to consider solutions 
of  (\ref{newtc0})  in detail. It is convenient to change
 the argument
$t\rightarrow x$  in such a way that
\[ r^2\frac{d}{dt} = - \frac{w}{2\sqrt { 2 \epsilon }}\frac{d}{dx}. \]
This is achieved for 
\begin{equation} \label{dx}
x = \frac{1}{2}\left[ \left(1+\frac{w}{\epsilon r}\right)^{1/2}+
1 \right].
\end{equation}
The new variable $x = x(\epsilon r)$ satisfies 
\begin{equation} \label{xas}
x\rightarrow \left\{ 
\begin{array}{ll}
\frac{1}{2}
\left( \frac{w}{\epsilon r}\right)^{1/2}& ~~~~ \epsilon r \ll 1 \\
1 +\frac{w}{4 \epsilon r}    & ~~~~\epsilon r \gg 1 
\end{array}
\right.
\end{equation}
Using  (\ref{dx}) one can rewrite  (\ref{newtc0}) as
\begin{equation} \label{hyp}
\frac{d^2 \chi(x)}{dx^2}= \frac{\lambda}{x(x-1)}\chi(x),
\end{equation}
where $\lambda$ is the coefficient 
\begin{equation} \label{lam}
\lambda = 2\frac{w''}{w}= 2 \frac{12 Z-1}{4 Z-1}.
\end{equation}
Scaling the function
\begin{equation} \label{xy}
\chi(x) \rightarrow y(x) = \chi(x)/x
\end{equation}
we find that the motion in the vicinity of the top of the ridge
is described by the following linear differential equation
\begin{equation} \label{dif}
x(1-x)y''(x) + 2 (1-x) y'(x)+\lambda y(x) =0,
\end{equation}
in which one  recognizes the known \cite{lan} hypergeometric equation
whose parameters $\alpha,\beta,\gamma$
\begin{equation} \label{abg}
\alpha=\nu+1,~~~\beta=-\nu,~~~\gamma=2
\end{equation}
are restricted by   the  condition on $\nu$
\begin{equation} \label{nunu}
\nu(\nu+1)=\lambda.
\end{equation}
Solving this equation and using (\ref{lam}) one finds an 
explicit form for this parameter  
\begin{equation} \label{nu}
\nu= \frac{1}{2}\left[ \left(\frac{100 Z-9}{4 Z-1}\right)^{1/2}-1\right],
\end{equation}
which governs the Wannier problem, compare  (\ref{nu}) with  (\ref{ex}).

It is convenient to choose a set of basic
solutions $y_1(x),y_2(x)$ of  (\ref{dif}) as 
\begin{eqnarray}\nonumber
y_1&=& (x-1)^{\gamma -\alpha-\beta}F\left(
\gamma-\alpha,\gamma-\beta,\gamma+1-\alpha-\beta,1-x\right),
\\ \nonumber
y_2& =& x^{-\alpha}F\left(\alpha,\alpha+1-\gamma,\alpha+1-\beta,1/x
\right),
\end{eqnarray}
where $F(\alpha,\beta,\gamma,x)$ is the  hypergeometric function.
Then an arbitrary solution of  (\ref{hyp}) 
can be presented as a linear combination
\begin{equation} \label{cfx}
\chi(x)= a_1 f(x) + a_2 g(x)
\end{equation}
of functions $f(x)=x y_1(x),~g(x)=x y_2(x)$
having the following form
\begin{eqnarray}\label{solchi}
f(x)&=& 
x(x-1)F\left(-\nu+1,\nu+2 ,2 ,1-x\right)
\\ \label{solchi2}
g(x)&=& 
\zeta \frac{\nu+1}{2 \nu+1}
x^{-\nu}F\left(\nu,\nu+1,2 \nu+2,\frac{1}{x}\right).
\end{eqnarray}
To simplify the subsequent formulae
a normalization coefficient $\zeta$ in  (\ref{solchi2})
is chosen to be
\begin{equation} \label{anu}
\zeta = \frac{[ \Gamma(\nu+1)]^2} { \Gamma(2 \nu+1)}.
\end{equation}
Here $\Gamma(x)$ is the usual Euler gamma function.
We will need to know the asymptotic conditions for the
functions  $f(x),g(x)$,  which are easily found from 
 (\ref{solchi}),(\ref{solchi2})
\begin{eqnarray}\label{asy}
f(x)\rightarrow 
\left\{ \begin{array}{cl}
\frac{1}{\zeta(\nu+1)}x^{\nu+1}
&  ~~~~~ x\rightarrow \infty \\
x-1
&  ~~~~~ x\rightarrow 1,
\end{array}
\right.
\end{eqnarray}

\begin{eqnarray}
\label{asy2}
g(x)\rightarrow \left\{
\begin{array}{cl}
\zeta\frac{\nu+1}{2\nu+1}
x^{-\nu} & ~~~~~ x\rightarrow \infty \\
1        & ~~~~~ x\rightarrow 1.
\end{array}
\right.
\end{eqnarray}
At the end points $x=1$ and $x=\infty$ one of the functions, either
$f(x)$ or $g(x)$, is singular while the other one is regular making them
 convenient for applications. 
Notice that the derivative $g'(x)$ remains  singular
at  $x\rightarrow 1$
\begin{equation} \label{sing}
g'(x)\rightarrow 
2 \frac{w''}{w}\left[
\ln (x-1)+d\right] + O[ (x-1) \ln (x-1) ].
\end{equation}
Here $d$ is a constant 
\begin{equation} \label{b}
d = 2\left[ \psi (\nu+1)+C \right],
\end{equation}
where
$\psi(x) =\Gamma'(x)/\Gamma(x)$ and
$C=-\psi(1)\simeq 0.577 $ is the Euler constant.
One more necessary parameter in the problem is 
the Wronskian of the functions $f(x),g(x)$
which can be found from 
asymptotic conditions (\ref{asy}),(\ref{asy2})
\begin{equation} \label{wron}
f'(x)g(x)-f(x)g'(x) =  1.
\end{equation}
The given consideration describes in detail  the main trajectory
along the top of the Wannier ridge 
 (\ref{r0}) as well as small deviations from it 
given in
 (\ref{cfx}),(\ref{solchi},(\ref{solchi2})
in which the variable $x = x(\epsilon r)$ 
is defined in  (\ref{dx}),(\ref{t(r)}).

\subsection{Quantum description}
Essential to the problem distances are in the Coulomb zone
\begin{equation} \label{er}
r \sim \frac{1}{\epsilon} \equiv r_{\rm c}.
\end{equation}
In this region the Coulomb force is 
$F_{\rm c}\sim 1/r^2 \sim \epsilon^2$.
The variation of the Coulomb energy on the 
wavelength $(1/\sqrt{\epsilon})F_{\rm c} \sim \epsilon^{3/2}$
proves to  be smaller than the kinetic energy $\sim \epsilon$ 
provided the energy is low 
\begin{equation} \label{eps}
\epsilon \ll 1. 
\end{equation}
It is supposed in this paper that the later condition  is valid.
The given estimations show that in this case
the semiclassical approximation turns out to be correct permitting one
to write  the wave function in the well known form
\begin{equation} \label{psi}
\psi_P(q) =N(q,P) \exp [ i  S(q,P)],
\end{equation}
where $ S(q,P) $ is the classical action.
The wave function depends on a set of coordinates $q$ and 
quantum numbers
$P$, the later ones play a role of parameters. It is convenient
to choose $P$ as momenta conjugate to coordinates $q$.
To be specific consider the convergent wave function for
the final state $\psi = \psi^{(-)}$ which is the most interesting in
the  Wannier problem.
This wave function 
may be looked at as a transition amplitude from 
the state in which coordinates $q$ of the Wannier pair are   
in the vicinity of the atom  to the final state in
which 
momenta $P$  describe the motion  at an infinite separation.
In classical terms  the transition 
should be described by a trajectory
with  initial coordinates $q$
and final momenta $P$. These 
two sets of parameters $q$ and $P$ fix the trajectory.
The classical action
depends on both these sets of variables $S =  S(q,P)$ .
It is well  
known that   the non-exponential factor $N(q,P)$ in the wave function
depends on the derivatives of the action \cite{fo}
\begin{equation} \label{N}
N(q,P) =
\left[ \det \left(-
\frac{\partial^2 S}{\partial q_i \partial P_j}\right) \right]^{1/2}.
\end{equation}
Let us choose a set of coordinates $q$
as $q=(r,\eta)$,
where $\eta$ is the momentum conjugate to the
variable $\chi$ introduced in  (\ref{r1})
\begin{equation} \label{pchi}
\eta = \frac{\partial L}{\partial \dot \chi} = r^2 \dot \chi.
\end{equation}
A convenience of this choice  becomes apparent soon.
The corresponding momenta are $P=(\bar p,\bar \chi)$.
(More accurately the momentum conjugate to $\eta$
should be written as $-\chi$. The  minus sign is dropped
to simplify notation.)
Here
$p$ is a momentum conjugate to $r$
\begin{equation} \label{pr}
p   = \frac{\partial L}{\partial \dot r} = \dot r.
\end{equation}
To avoid confusion 
the bar symbol is used to mark the variables
belonging to the final point of the trajectory for infinite separations.
In this  notation the energy of the pair is written 
as $\epsilon = \bar p^2/2$, 
and the action looks like
$S = S( r, \eta; \bar p,\bar \chi)$.
Calculating the action we should consider the  region in the
vicinity of the Wannier ridge where both $\bar \chi$ and $\eta$ are 
small.
Therefore we can expand the  action in powers of these variables
\begin{equation} \label{Del}
S( r, \eta; \bar p,\bar \chi) = \hat S_0(r,\bar p) + 
\hat \Sigma(r,\eta;\bar p,\bar \chi),
\end{equation}
where 
\begin{equation} \label{s0}
\hat S_0(r,\bar p)  = S( r, 0; \bar p,0)=\int_r^{\bar r} 
\sqrt{ \bar p^2 + 2 \frac{w}{r'} }dr' - \bar p \bar r
\end{equation}
describes a one-dimensional motion along the ridge-top trajectory
governed by the effective Coulomb charge $w$.
We will recognize later, see  (\ref{rac}),
that a slightly different choice of the effective 
charge  proves to be more convenient.
The quantities which will be recognized be affected by this  
redefinition are marked by the hat symbol.
The quantity $ \hat \Sigma(r,\eta;\bar p,\bar \chi)$ 
depends on the derivatives of $S$. 
The first derivatives for the ridge-top trajectory are obviously zero.
Therefore, the  expansion  
starts from the second powers of $\bar \chi,\eta $. 
It is shown  below, see discussion after  (\ref{answ}),
that it is sufficient to restrict  our consideration
to these lowest order terms writing
\begin{equation} \label{s2}
\hat \Sigma(r,\eta;\bar p,\bar \chi) = 
 \frac{\bar \chi^2}{2\bar p}\hat A+ 
\bar \chi \eta B +  
 \frac{\eta^2\bar p}{2}C.
\end{equation}
The factors $1/\bar p$ and $\bar p$ are included here 
to simplify the subsequent formulae.
The coefficients of this expansion are functions of $r$ and $\bar p$.
In order to find them  let us remember that
 $\chi$ and $\eta$ are the conjugate variables and therefore
\begin{equation} \label{chi2}
\frac{1}{\bar p}\hat A =   \frac{\partial^2 S}{\partial \bar\chi^2}=
\left( \frac{\partial \bar \eta}{\partial \bar \chi}\right)_{\eta},
\end{equation}

\begin{equation} \label{chip}
 B = \frac{\partial^2 S}{\partial \bar \chi \partial \eta}=
\left( \frac{\partial \bar \eta}{\partial \eta}\right)_{\bar\chi},
\end{equation}

\begin{equation} \label{p2}
\bar p C =  \frac{\partial^2 S}{\partial \eta^2}=
\left( \frac{\partial \chi}{ \partial \eta}\right)_{\bar \chi}.
\end{equation}
The derivatives in  (\ref{chi2}),(\ref{chip}),(\ref{p2})
describe the variation of the  coordinate and momentum under 
variation of the
trajectory. This variation in turn is described by  variation of the  
coefficients $a_1,a_2$ in  (\ref{cfx}) which governs the coordinate
$\chi(x) = a_1 f(x)+a_2 g (x)$ as well as the momentum $\eta$
conjugate to $\chi$
\begin{equation} \label{pchiaa}
\eta = r^2 \dot \chi = - \frac{w}{2 \bar p}\chi'(x)=
-\frac{w}{2\bar p}(a_1 f'(x)+a_2 g'(x)).
\end{equation}
Using  (\ref{cfx}),(\ref{pchiaa}) one finds from  (\ref{chi2})
\[ \hat A =
-\frac{w}{2}\left[
\frac {(da_1) \bar f' + (da_2) \bar g'}{(da_1) \bar f + (da_2) \bar g}
\right]_\eta \]
where according to  (\ref{pchiaa})
the condition $\eta = const$ restricts  variations 
of coefficients $da_1,da_2$
\[ (da_1) f' + (da_2) g' = 0.\]
Combining the last two equations we find 
\begin{equation} \label{Achi2}
\hat A = 
\frac{w}{2}\frac{ \bar f'g'- f'\bar g'}{f' \bar g-\bar fg'}.
\end{equation}
The other derivatives are calculated similarly
\begin{equation} \label{B}
 B =  
\frac{ f'g - f g'}{f' \bar g-\bar fg'}=\frac{1}{f' \bar g-\bar fg'},
\end{equation}

\begin{equation} \label{C}	
C = 
-\frac{2}{w}\frac{ f \bar g-\bar f g }{f' \bar g-\bar fg'}.
\end{equation}
Evaluating the last identity in  (\ref{B}) the Wronskian  (\ref{wron}) 
was used.
All three coefficients  turn out  to be   functions of the single 
variable $x$, $A=A(x),B=B(x),C=C(x)$.
This property of the coefficients was anticipated when normalization 
factors  $1/\bar p$ and $\bar p$ were introduced in 
their definitions  in  (\ref{s2}).

We can now use  the fact that for the final state $\bar r = \infty$,
and therefore according to
 (\ref{xas})  $\bar x= 1$.
From   (\ref{asy}),(\ref{asy2})  we find
$ \bar f = 0,~~~\bar f'=1,~~~\bar g = 1$, resulting in
\begin{eqnarray}\label{S2si}
\hat A(x)
&=&\frac{w}{2}\left( \frac{g'(x)}{f'(x)}-\bar g'
\right), \\ \label{S2si2}
B(x)
& =& \frac{1}{f'(x)},\\ \label{S2si3}
C(x)
 &=&-\frac{2}{w}\frac{f(x)}{f'(x)}.
\end{eqnarray}
All these quantities are well defined functions of $x$,
except for  the constant $\bar g'$ which
according to  (\ref{sing}) is logarithmically
divergent $ \bar g' = g'(\bar x)\sim \ln(\bar x-1),~~\bar x \rightarrow
\infty$.
This divergence  manifests itself in the action as a term
\begin{equation} \label{ln}
-\frac{w\bar\chi^2}{4 \bar p}g'(\bar x)
=
\frac {w'' \bar\chi^2}{2\bar p}
\left[ \ln \left(\frac{2 \bar p^2 \bar r}{w}\right) - d \right].
\end{equation}
In order to deal with this problem let us remember that
the Coulomb action $\hat S_0$   (\ref{s0})
reveals the logarithmic divergence as well,
$\hat S_0 \sim (w/\bar p) \ln (2 \bar p^2 \bar r/w), ~~\bar r \rightarrow
\infty$. This later divergence proves   harmless.
Really, it takes place for an effective one-dimensional Coulomb problem
described by the only coordinate $r$. For this simple situation
the Coulomb logarithmic divergence is of course well studied and
cured by textbook methods, see \cite{lan}. 
Having this in mind let us  ``renormalize''
$\hat S_0$ and $\hat \Sigma$, $\hat S_0\rightarrow S_0,~\hat \Sigma
\rightarrow \Sigma$
\begin{equation} \label{sr}
S(r,\eta;\bar p,\bar \chi) = S_0(r,\bar p) + 
\Sigma(r,\eta;\bar p,\bar \chi)
\end{equation}
in a way that makes
the divergence  (\ref{ln}) apparently harmless as well.
The  action $S_0(r,\bar p)$ in  (\ref{sr}) is defined as
\begin{equation} \label{rac}
 S_0(r,\bar p) = 
\int_r^{\bar r} \sqrt{ \bar p^2 + 2 \frac{w(\bar \chi)}{r'} } 
dr' - \bar p \bar r. 
\end{equation}
It describes the one-dimensional motion in the  field of 
the Coulomb charge $w(\bar \chi)$  (\ref{W}) which explicitly depends
on $\bar \chi^2$, see  (\ref{w}). This property distinguishes it from
the ``non-renormalized'' 
quantity $\hat S_0$  (\ref{s0}) which does not depend
on $\bar \chi$.
The renormalized  $\Sigma$ 
\begin{equation} \label{dsr}
\Sigma(r,\eta;\bar p,\bar \chi) = \hat \Sigma(r,\eta;\bar p,\bar \chi)  
- \frac{w''\bar \chi^2}{2}\int_r^{\bar r}
\frac{dr'}{ r'\sqrt{ \bar p^2 +2 w/r' }}
\end{equation}
is defined in such a way
that it compensates for the dependence of $S_0$
on $\bar \chi^2$.
Using these new definitions we find
\begin{equation} \label{abc}
\Sigma(r,\eta;\bar p,\bar \chi) = 
\frac{\bar \chi^2}{2\bar p} A(x) +  \bar \chi \eta B(x)
+ \frac{\eta^2\bar p}{2}C(x),
\end{equation}
where $B(x),C(x)$  remain
equal to definitions given in (\ref{S2si2}),(\ref{S2si3}). In contrast
$A(x)$ 
differs from the ``non-renormalized'' value $\hat A(x)$ (\ref{S2si}),
namely
\begin{equation} \label{are}
A(x) =  
\frac{w}{2 } \frac{ g'(x)}{f'(x)}+w''h(x) .
\end{equation}
Here $h(x)$ is a new function  whose definition follows from
 (\ref{ln}),(\ref{dsr})
\[
h(x) = 
-  \bar p\int_r^{ \bar r}\frac{dr'}{r' \sqrt{\bar p^2 + 2 w/r'}}
+ \ln\left( \frac{2 \bar p^2 \bar r}{w} \right) -d.
\]
Calculating the integral and taking the limit $\bar r \rightarrow \infty$
we find
\begin{equation}\label{h}
h(x) = 
\ln \left[ \frac{r}{w} \left( \bar p^2 +\frac{w}{r} + 
\bar p \sqrt{  \bar p^2 + \frac{2 w}{r} } \right)  \right]  - d 
= \ln \frac{x}{x-1}-d,
\end{equation}
where $d$ was defined in  (\ref{b}).

The non-exponential factor $N(q,P)$ in the wave function
 (\ref{psi}) is easily calculated
because with the necessary accuracy 
it is
\begin{equation} \label{n1n2}
N(r,\eta; \bar p,\bar \chi) 
\simeq N_0 (r,\bar p)\left( \frac{\partial^2 S}
{\partial \bar \chi \partial \bar \eta}\right)^{1/2}=N_0 (r,\bar p) B(x).
\end{equation}
Here the first factor $N_0(r,\bar p)$ is related to the
effective  one-dimensional Coulomb
problem
\begin{equation} \label{AA}
N_0(r,\bar p) = 
\left( -
\frac{\partial^2 S_0}{\partial r \partial \bar p} 
\right)^{1/2}.
\end{equation}
The factor $B(x)$ 
was calculated previously in  (\ref{S2si2}).

Combining the above calculated quantities 
we can present the  wave function  (\ref{psi}) in the 
following final form
\begin{equation} \label{answ}
\psi_{p,\chi}(r, \eta)=
\sqrt{B(x)} \exp [ i \Sigma(r,\eta;p,\chi) ]
\phi^{(-)}_p (r; w(\chi)).
\end{equation}
We need no more to distinguish by a bar symbol the variables
at the final point  supposing that
$p \equiv \bar p,~\chi \equiv \bar \chi$.
Evaluating  (\ref{answ}) it was taken into 
account that $N_0(r) \exp (i S_0)$
gives the semiclassical representation for the  
Coulomb wave function $\phi^{(-)} (r,p; w( \chi))$
\begin{equation} \label{sem}
 \phi^{(-)} (r,p; w( \chi)) = N_0(r) \exp (i S_0).
\end{equation}
This single-particle wave function
gives an effective description of propagation 
of the Wannier pair near the top 
of the Wannier ridge. The propagation is characterized by 
the momentum $ p$ and the effective Coulomb charge $w( \chi)$.
It  is often more convenient to work
directly with the Coulomb wave function rather than with its
semiclassical representation  (\ref{sem}), though one can keep open a
possibility to return to the semiclassical description when necessary.

The quantity
$ \Sigma(r,\eta;p,\chi) $  in  (\ref{answ}) is defined by expansion
 (\ref{abc}), the coefficients 
$A(x),B(x),C(x)$
of which are given in  (\ref{are}),(\ref{S2si2}) and (\ref{S2si3}).
From  (\ref{answ})  we conclude that
typical values  for both $ \chi$ and $\eta$  are restricted
by the available above-threshold energy. For $ \chi$
we find   $  \chi^2\sim  p$. Similarly, for
$\eta$ we find  that
$\eta^2 \sim  p \eta_0^2$. Here $\eta_0$  is a typical
classical value of the momentum  $\eta$ which according to  
(\ref{pchiaa})
is estimated as $\eta_0=const/ p$.
For low  energy, see   (\ref{eps}),  both $ \chi$ and 
$\eta$ are small, $ \chi^2 \ll 1, ~\eta^2 \ll \eta_0^2$. This fact
justifies expansion  (\ref{abc}) of the action in powers of $ \chi$ and
$\eta$.

Along with the wave function  (\ref{answ}) one can consider
another solution for the Wannier pair
$\sim \sqrt{B} \exp(-i \Sigma) \phi^{(+)}$.
We use conventional notation which shows explicitly the asymptotic
behaviour of the Coulomb wave functions
\begin{equation} \label{ca}
\phi^{(\pm)}_p(r; w( \chi)) 
\rightarrow \frac{1}{ \sqrt{p}} 
\exp \left[ \pm i\left( p r +\frac{w(\chi)}{ p}
\ln 2  p r    \right) \right].
\end{equation}
The normalization coefficient $1/\sqrt{p}$ is convenient
for the following calculations, see   (\ref{Ll}) in 
Section \ref{conseq}.
At this stage  an advantage  of representation
in which $ \chi$ is a parameter and $\eta$ is an argument 
of wave functions  becomes obvious. 
The potential 
$U(r,\chi)$  (\ref{U}) approaches the value $-w( \chi)/r$ for
$r\rightarrow \infty$ resulting in the effective Coulomb
charge $w( \chi)$
which in this representation does not depend on arguments of  
the  wave function
and may be considered as a given constant.
We have used this fact above to satisfy  boundary conditions 
for large separation. In other representations, for example
in representation in which
$\chi$ is an argument of a wave function and $\eta$ is a parameter
it is more laborious to satisfy these boundary conditions.

For effective use of the wave function  (\ref{answ})
we will need to know the asymptotic form of  the coefficients
$A(x),B(x),C(x)$  in  (\ref{abc}).
It can be found from asymptotic conditions 
 (\ref{asy}),(\ref{asy2}).
Inside the Coulomb zone $\epsilon r \ll 1$,
we find 
\begin{eqnarray}
   \label{aScc}
&& A(x)
\simeq  
-w''d,
\\ \label{aSpc}
&&B(x)
\simeq  \zeta x^{-\nu}
=  \zeta \left(\frac{ 2 p^2 r}{w} \right)^{\nu/2},
\\ \label{aSp}
&&C(x)
\simeq  - \frac{1}{\nu+1}\frac{2 }{w}x =
 - \frac{1}{\nu+1} \left( \frac{2 }{wp^2 r}\right)^{1/2}.
\end{eqnarray}
Outside the Coulomb zone $\epsilon r \gg 1$ we find
\begin{eqnarray}\label{11}
A(x)
& \simeq&  -\frac{2 w''^2} {p^2 r} \rightarrow 0,\\ \label{12}
B(x)
& \simeq& 1, \\ \label{22}
C(x)
&\simeq& - \frac{1}{ p^2 r}\rightarrow 0.
\end{eqnarray}
Using  (\ref{11}),(\ref{12}),(\ref{22}) we verify that
the phase in the definition of the wave function  (\ref{answ}) is chosen 
in such a way that  for $r \rightarrow \infty$
\begin{equation} \label{pas} 
\psi_{p,\chi}(r,\eta) \rightarrow \frac{1}{\sqrt{p}}
\exp \left[ i\left(  \eta \chi -
pr - \frac{ w(\chi)}{p} \ln 2 pr\right) \right].
\end{equation}

The  discussion above shows that
the main features of the Wannier problem are described by 
the effective single-particle  Coulomb problem with
the effective Coulomb charge $w(\chi)$.
An essential  property of the
single-particle problem is a phase shift. 
In order to find it let us present it in the following form
\begin{equation} \label{ps}
\delta(\chi) = \delta_{\rm c}(\chi) +\Delta \delta (\chi).
\end{equation}
Here the first term 
is the usual Coulomb phase shift 
which arises due to the effective motion in the
Coulomb field of the charge $w(\chi)$
\begin{equation} \label{cp}
\delta_{\rm c}(\chi) = \arg \Gamma\left( 1-i \frac{w(\chi)}{p}\right).
\end{equation} 
An additional phase shift $\Delta \delta(\chi)$ is produced by the 
term $\Sigma(r,\eta;p,\chi)$ in the exponent
in  (\ref{answ}). 
Remember that this term is part of the classical action.
The variation of the action when $r$ varies from $r=\infty$ 
to $r = 0$ in the semiclassical  approximation is
identical to the phase shift. Therefore 
\begin{equation}\label{Dd}
\fl \Delta \delta (\chi) = -\left[ 
\Sigma(r=\infty,\eta=0;p,\chi) -\Sigma(r=0,\eta=0;p,\chi)\right ] 
=- \frac{w''d}{2p}\chi^2.
\end{equation}
Evaluating the last equality  (\ref{aScc}),(\ref{11}) were used.
The minus sign in front of the square brackets in  (\ref{Dd})
agrees with  definition  (\ref{ca}) of the function 
$\phi^{(-)}_p(r;w(\chi))$.
From the asymptotic condition  (\ref{pas}) we conclude that the found
phases  (\ref{cp}),(\ref{Dd}) are the only ones which contribute to
the phase shift  (\ref{ps}).

The found representation for the wave function of the Wannier problem
 (\ref{answ}) has a clear physical meaning.
The Wannier
problem takes place in the vicinity of the Wannier ridge.
This allows to present
essential properties of the
two-electron  problem  by the effective
single-particle Coulomb problem.
The Coulomb wave function $\phi^{(-)}_p(r;w(\chi))$
of this effective Coulomb problem 
gives the simplest approximation.
The nontrivial three-body nature of the Wannier problem
is taken care of  by
the quantities $\Sigma(r,\eta;p,\chi),B(x)$ which accounts for 
small fluctuations in the vicinity
of the ridge-top trajectory.
These fluctuations result in two interesting properties.
Firstly,
the state of the Wannier pair on the top of the ridge is unstable
in respect to sliding down into configuration when one electron is 
close to
the atom while another one is far apart.
This reduces the probability for the pair
to survive on the top of the ridge. This effect is
taken into account by the normalization coefficient $B(x)\le 1$.
Secondly, there appears  an additional important contribution to the 
phase shift $\Delta \delta(\chi)$ in the effective single-particle
problem.

If the  wave function  (\ref{answ}) is used in the traditional Wannier 
problem to calculate the cross-section 
of a process with two Wannier electrons in the final
state  then the normalization coefficient $B(x)$ 
should be estimated
with the help of  (\ref{aSpc}) in the vicinity of the atom. This gives 
$|\psi| \sim  \varepsilon^{\nu/4}$, and 
therefore 
the threshold behaviour of
the  cross-section is estimated as
 $\sigma \sim |\psi|^2 \sim \varepsilon^{\nu/2}=\varepsilon^\mu $
reproducing the Wannier threshold law  (\ref{w1}).

\section{PCI in the near-threshold region}
\label{near}
The resonant nature of reaction 
 (\ref{1}),(\ref{2}) permits one to
present the matrix element describing it in the following general form
\begin{equation} \label{me}
M = \sum_{\rm j}\
\frac{ \langle \Psi_{\rm f} | V_{\rm A} |\Psi_{\rm j} \rangle
\langle \Psi_{\rm j} | V_{\rm col}| \Psi_{\rm i}\rangle  }
{\varepsilon - \varepsilon_{\alpha {\rm j}} -\varepsilon_{\beta {\rm j}} 
+ i\Gamma/2}.
\end{equation}
The wave function $|\Psi_{\rm i}\rangle$ describes  the initial state 
${\rm e + A}$,
$ |\Psi_{\rm j} \rangle$ describes the set of intermediate 
states ${\rm e_\alpha+e_\beta+A}^{+*}$  with the fixed resonant state
of the ion ${\rm A}^{+*}$, 
and $|\Psi_{\rm f}\rangle$ describes the final
state ${\rm e_\alpha+ e_\beta+ e_\gamma+A}^{2+}$ 
of the reaction described by  (\ref{1}),(\ref{2}).
Summation in  (\ref{me}) takes into account all possible
states for Wannier
electrons ${\rm e_\alpha,e_\beta}$ in the intermediate state.
The potential $V_{\rm col}$ describes the part of the Hamiltonian 
which is
responsible for the collision, and $V_{\rm A}$ is the potential 
which causes
the Auger decay. The energy denominator depends on the above-threshold
energy as well as on the 
energies $\varepsilon_{\alpha {\rm j}},\varepsilon_{\beta {\rm j}}$ 
of the Wannier electrons in the intermediate state
and the total width $\Gamma$ of the autoionizing atomic state.
We are interested in the low near-threshold energies for the Wannier pair
 (\ref{eps}). In contrast the autoionized electron is sufficiently 
fast and 
we can assume that its energy $\varepsilon_\gamma$ is so high that
\begin{equation} \label{3>}
\varepsilon \ll 1\ll \varepsilon_\gamma. 
\end{equation}
The width $\Gamma$
is another small parameter of the problem.
We will assume that the width is much lower than the above threshold 
energy
\begin{equation} \label{ge}
\Gamma \ll \varepsilon.
\end{equation}
Inequalities  (\ref{3>}),(\ref{ge}) 
greatly simplify the problem. 
Firstly,   (\ref{ge}) implies that
the decay takes place when the Wannier pair
is far 
outside the atomic radius. As a result the electrons ${\rm e_\alpha,
e_\beta}$ have 
no influence on the process of ejecting the electron ${\rm e_\gamma}$.
Secondly, inequality  (\ref{3>}) guarantees that the 
interaction between the Wannier pair and the autoionized electron
is small, and we can neglect it. As a result the matrix element
$\langle \Psi_{\rm f} | V_{\rm A} |\Psi_{\rm j} \rangle$ reduces to
\begin{equation} \label{ad}
\langle \Psi_{\rm f} | V_{\rm A} |\Psi_{\rm j} \rangle
\simeq \sqrt {\Gamma_{\rm f} }
\langle \psi_2 |\psi_{\rm j}\rangle.
\end{equation}
Here  $\Gamma_{\rm f}$
is the partial width of the autoionizing state, which makes
$ \sqrt{ \Gamma_{\rm f}} $  proportional to
the matrix element responsible for the
autoionization decay. 
It is determined by the processes which take place 
strictly inside the atomic particle and,
as  mentioned,  are not influenced by PCI.
The effects of PCI are taken care of in the overlapping integrals
$\langle \psi_2 |\psi_{\rm j}\rangle$
between the wave functions which describe the states of the Wannier pair.
We denote by $\langle \psi_2|$ the final-state wave-function 
for the Wannier pair  and
by $|\psi_{\rm j}\rangle $ 
the full set of the intermediate-state wave functions for the pair.
Substituting  (\ref{ad}) in  (\ref{me}) we can present 
the matrix element $M$ in the following simple form 
\begin{equation} \label{gr}
M = \sqrt{ \Gamma_{\rm f}} \langle \psi_2 | \psi_1\rangle.
\end{equation}
Here the wave function $|\psi_1\rangle $ is defined as
\begin{equation} \label{psi0}
|\psi_1\rangle =  \langle 
\Psi ({\rm A}^{+*})  | 
G^{(+)}_{\varepsilon+i\Gamma/2}|V_{\rm col}|\Psi_i\rangle,
\end{equation}
where $\langle  \Psi ({\rm A}^{+*})|$ 
is the wave function of the resonant state of the ion.
It projects the many-electron wave function $V_{\rm col}|\Psi_i\rangle$
onto the states of the Wannier electrons.
The function $|\psi_1\rangle$ depends on the 
coordinates of the two electrons.
Summation over the full set  
of intermediate states for the pair is
included in the two-particle Green function
\begin{equation} \label{gf}
G^{(+)}_{\varepsilon+i\Gamma/2} \equiv
G^{(+)}_{\varepsilon+i\Gamma/2}({\bf r}_1,{\bf r}_2;
{\bf r}_1',{\bf r}_2')=
\sum _{\rm j}
\frac{\psi_{\rm j}({\bf r}_1,{\bf r}_2) 
\psi^*_{\rm j}({\bf r}_1',{\bf r}_2')}{\varepsilon - 
\varepsilon_{\alpha {\rm j}}-\varepsilon_{ \beta {\rm j}} 
+ i\Gamma/2}.
\end{equation}
Equation (\ref{gr}) states that the influence of PCI  on the Wannier
pair can be described with the help of the overlapping integral.
This result 
is very close in spirit to the shake-down model 
\cite{read2,shd}  
which originally was developed for the case when one near-threshold 
electron 
is exposed to the variation of the charge of the atomic particle caused 
by the autoionization. In the case under consideration there are two
near-threshold electrons, but the idea that overlapping of the wave
functions gives the correct description of PCI remains the same. 

The wave functions in the matrix element  (\ref{gr}) satisfy the 
Schroedinger equations
\begin{eqnarray}\label{sch1}
\left( \varepsilon +i\frac{\Gamma}{2}- H_1\right)
| \psi_1\rangle &=& |Q\rangle,  \\ \label{sch2}
( E - H_2)| \psi_2\rangle &=& 0. 
\end{eqnarray}
Here $H_1$ and $H_2$ are the Hamiltonians which describe propagation
of the Wannier pair in the intermediate and final states
\begin{equation} \label{ha}
H_n = -	\frac{1}{2}(\Delta_\alpha+ \Delta_\beta)+
\frac{1}{r_{\alpha\beta}}
-Z_n\left(\frac{1}{r_\alpha}+\frac{1}{r_\beta}\right).
\end{equation}
An index $n=1,2$ here labels the intermediate and the final states
in which 
the Coulomb charge of the atomic particle
is  different.
Generally speaking there are also the short-range
potentials which account for the 
potential of the atomic particle in the intermediate or final states.
However,  the influence of the 
short-range potential on PCI is small, and these potentials are 
neglected in
 (\ref{ha}) and  the following consideration.
The energy 
$\varepsilon+i\Gamma/2$ of the pair in the state $|\psi_1\rangle$
is complex. Its real part $\varepsilon$ is equal to the above-threshold
energy, its imaginary part $\Gamma/2$ appears due to autoionization
decay. The function $|\psi_2 \rangle$ describes the Wannier pair
in the final state with real energy $E$ which satisfies the energy
conservation law $E_{\rm col} + E({\rm A}) =
E + \varepsilon_\gamma + E( {\rm A}^{2+})
$, where $E({\rm A})$ is the energy of the atom,
$ E( {\rm A}^{2+})$ is the energy of the ion, $\varepsilon_\gamma$
is the energy of the autoionized electron, and $E_{\rm col}$
is the collision energy.

According to  (\ref{psi0}),(\ref{gf}) there appears a source
in the right-hand side of the Schroedinger equation (\ref{sch1})
\begin{equation} \label{q}
| Q \rangle =\langle \Psi({\rm A}^{+*}) | V_{\rm col}| \Psi_i\rangle
\end{equation}
which is a function of the coordinates of the pair.
For large separations from the ion this function becomes proportional
to the amplitude of inelastic scattering and inversely proportional
to separation.
It is important that the effects of PCI take place far outside
the atomic particle \cite{ks}, 
where the source is   small
$\langle {r}_\alpha,{r}_\beta | Q \rangle \simeq 0,~~r_\alpha,
r_\beta \gg1$.
Therefore
it is sufficient for our purposes to solve  only the
homogeneous Schroedinger equation for the function
$|\psi_1\rangle $
\begin{equation} \label{hs}
\left(\varepsilon +i \frac{\Gamma}{2}- H_1 \right) |\psi_1\rangle = 0
\end{equation}
in the region far outside  the atom and then 
normalize this function  to make it  proportional
to the amplitude of inelastic scattering, thus taking into account
the influence of the source.

The given consideration shows that PCI for the Wannier pair in
the vicinity of the threshold is described by the 
overlapping integral of the wave functions describing the pair
in the intermediate and final states  (\ref{gr}).
Notice that this possibility   follows from 
conditions  (\ref{3>}),(\ref{ge}).
In this sense it is rather a theoretical {\it result} then a {\it model}.

\section{PCI for the Wannier pair}
\label{pci}
Let us calculate the overlapping integral 
 (\ref{gr}) using 
the wave functions for the Wannier problem evaluated in 
Section \ref{wan}.
The model suggested  in Section \ref{wan} 
describes the state of the Wannier  pair
by its energy and the  quantum number $ \chi$.
The amplitude of the inelastic collision 
depends on these quantum numbers
\begin{equation} \label{ain}
{\cal A} ( \chi) = \langle \varepsilon, \chi; 
\Psi({\rm A}^{+*})| V_{\rm col}|\Psi_i\rangle.
\end{equation}
The matrix element 
in the right-hand side
describes the events which take place in the vicinity
of the atom.
The small width $\Gamma$ 
of the autoionizing state
does not manifest itself for these small separations.
That is why we can neglect it, supposing
that the wave function  $\langle \varepsilon, \chi |$ in  (\ref{ain})
describes  a real physical state of the pair with the real energy 
$\varepsilon$.

The solution of  (\ref{hs}) can be presented in the following form
\begin{equation} \label{so}
|\psi_1 \rangle =  \int
\left[\psi_{ k, \chi'} (r,\eta)\right ] _ 1 {\cal A}( \chi')  d \chi'.
\end{equation}
Here the wave function $\psi_{ k, \chi'} (r,\eta)$
defined in  (\ref{answ}) describes a state of the Wannier pair
\begin{equation}\label{iwf}
\left[ \psi_{ k, \chi'} (r,\eta)\right]_1 = 
 \left[\sqrt{B(x)} \exp[ - i \Sigma(r,\eta;k,\chi')]
 \phi^{(+)}_k(r;w(\chi')) \right]_1.
\end{equation}
The  square brackets with the index 1 are
used  to remind that the Wannier pair is in the intermediate state
in which the Coulomb charge of the ion is $Z=Z_1=1$.
Asymptotic condition  (\ref{pas}) guarantees that the coefficients
${\cal A}(\chi')$ in  (\ref{so}) are identical to the amplitude
of inelastic scattering. 

It is important that
dependence of the amplitude ${\cal A}(\chi')$
on $\chi'$ can be established explicitly.
Remember that  (\ref{answ}) expresses the properties of the Wannier 
problem 
in terms of the effective single-particle problem. 
This allows one to present
the amplitude of the collision 
as the following  product
\begin{equation} \label{sp}
{\cal A}(\chi') = \exp [i \delta(\chi')] {\cal A}_0(\chi') \simeq
 \exp [i \delta(\chi')] {\cal A}_0(\chi) .
\end{equation}
Here $\delta(\chi')$ is the phase shift which was found
in   (\ref{ps}),(\ref{cp}),(\ref{Dd}). The quantity 
${\cal A}_0(\chi')$ 
is a matrix element of inelastic collision. 
Representation  (\ref{sp}) has a form which is conventional in
the processes with single-particle final states \cite{lan}.
The matrix element ${\cal A}_0(\chi')$
originates from  integration in the vicinity
of the atom.
In contrast,  the phase shift
$\delta(\chi)$ arises due to  the events which
take place far outside  the atom in the Coulomb zone.
Thus the  two quantities ${\cal A}_0(\chi')$ and $\delta(\chi')$
originate from quite different distances. This  makes  them
depend differently on $\chi'$. The phase shift
exhibits strong variation $\delta(\chi') \sim \chi'^2/k$, 
while the amplitude
${\cal A}_0(\chi')$  varies smoothly ${\cal A}_0(\chi') \sim 
 \chi'^2$.
In order to verify this statement one should  remember
that in the semiclassical region all quantities depend on 
the classical action
$\sim \int dr \sqrt{k^2+w(\chi')/r}$. 
For large $r$, $r \gg k^2$, the action
	has a term $\sim \chi'^2/k$
which  strongly varies with $\chi'$,
 while for small $r$, $r \ll k^2$,
there  is only a smooth behaviour $\sim\chi'^2\ll 1$.
More general verification of this property can be fulfilled
using the Landau-Smorodinsky approach
to the  problem of   a low-energy particle 
influenced by  a  long-range Coulomb
field as well as a short-range potential. It is discussed in detail 
in  \cite{lan}.

The smooth  behaviour  of the matrix element 
${\cal A}_0(\chi')$
with $\chi'$ justifies the last equality in  (\ref{sp}).
Using it we simplify  (\ref{so})
\begin{equation} \label{so1}
|\psi_1 \rangle = {\cal A}_0(\chi)
\int
\left[ \exp [i\delta (\chi') ]
\psi_{ k, \chi'} (r,\eta)\right ] _ 1   d \chi'.
\end{equation}
The momentum $k$ in the state  $|\psi_1 \rangle$
is a complex number due to the finite width 
of the autoionizing state
\begin{equation} \label{peg}
k  = \sqrt{ 2 \varepsilon + i \Gamma} \simeq \sqrt{2 \varepsilon} + 
i\frac{\Gamma}{2\sqrt{2\varepsilon}}.
\end{equation}
Inequality  (\ref{ge}) permits one to make expansion in the last equality
in  (\ref{peg}).
It is important to keep the width in this expansion because
the wave function  (\ref{so}) will be used to calculate the effects
of PCI which take place
for large separations. The  positive sign of the imaginary part of $ k$
ensures  that the wave function decreases exponentially
$ [\psi_{ k, \chi} (r,\eta)]_1 \sim \exp(-{\rm Im}\,   k r)$
with separation. 
Thus the events which  result in  PCI are localized in the finite region.
Similar localization of  the wave function
is well known  \cite{ks}
for the case when one electron is influenced by PCI. We see that 
localization
in the vicinity of the atom
\begin{equation} \label{lo}
r\le  
  \frac{1}{ {\rm Im}\,k } \simeq \frac{2\sqrt{2\varepsilon} }
{ \Gamma } \equiv
r_{\rm l}
\end{equation}
remains true for the Wannier pair as well.
The radius of localization $r_{\rm l}$ has a clear physical meaning
being proportional to 
classical distances $r_1(t) \simeq r_2(t) \simeq  r_{\rm l}/(2\sqrt{2})$ 
which separate Wannier electrons from the atom 
 at the moment $t=1/\Gamma$ of the decay.

The wave function $\langle \psi_2|$ in the overlapping integral
 (\ref{gr}) describes the real final
state of the Wannier pair with the real energy $E$. 
Generally speaking this state should be described by a linear 
combination
of outgoing and incoming waves. For our purposes
it is necessary to take into account the convergent wave 
because it gives the major contribution to the overlapping integral
and therefore correctly describes PCI. 
Really, the  convergent wave function behaves 
like $\sim \exp (-i pr),~p=\sqrt{2 \varepsilon}$
therefore the overlapping integral with the divergent
wave function $\sim \exp (ikr)$
can be roughly estimated as $\sim 1/(k-p)\sim 1/\Gamma$ 
describing the resonance. In contrast, the divergent wave
does not contribute to the resonance.
This shows that  the final-state wave function $|\psi_2 \rangle$ 
can be taken
in the form  (\ref{answ})
\begin{equation}\label{fswf}
| \psi_2 \rangle = \left[ \psi_{ p, \chi} (r,\eta) \right]_2 = 
 \left[\sqrt{B(x)} \exp[ i \Sigma(r,\eta;p,\chi)] 
 \phi^{(-)}_p(r;w(\chi)) \right]_2.
\end{equation}
The square brackets with the index 2 are to remind  that
all the quantities in the square brackets should be calculated
for the final state in which
the Coulomb charge of the ion is $Z=Z_2 = 2$ and the momentum of the 
pair is
$ p = \sqrt{2 E}$.
Notation introduced  in  (\ref{iwf}),(\ref{fswf}) will be 
used below as well. All quantities labeled by an index 1 are considered
for the intermediate state of the Wannier pair.
In this state the ion charge is $Z=Z_1 = 1$, 
the energy of the pair is
$\varepsilon+ i\Gamma/2$, the corresponding momentum is $k$  (\ref{peg}).
The quantities labeled by an index 2 describe the final state
with $Z=Z_2 = 2$, the final-state energy of the pair $E$ and 
the corresponding momentum $ p = \sqrt{2 E}$.

The explicit form of the wave functions 
 (\ref{so1},(\ref{fswf})
permits to  present  the matrix element
describing PCI  (\ref{gr}) in the following final form
\begin{equation}\label{ove}
M =
\sqrt{ \Gamma_{\rm f} }{\cal A}_0(\chi) S_{21},
\end{equation} 
where
\begin{equation}\label{2|1} 
S_{21} = \int   
[ \psi_{p,\chi}(r,\eta)]_2 
[\psi_{k,\chi'}(r,\eta)]_1 \exp [i \delta_1 (\chi') ]
\frac{d\chi'd\eta dr}{2\pi}.
\end{equation}
The above given consideration shows that this representation 
has a simple physical meaning. The fast autoionization decay
results in an abrupt  transition of the Wannier pair 
from the intermediate state to the final state. This transition
is described by the overlapping integral
of the wave functions for the intermediate and final states
\begin{equation} \label{over}
\langle \psi_{p,\chi,2} | \psi_{k,\chi',1} \rangle =  \int
[ \psi_{p,\chi}(r,\eta)]_2 
[\psi_{k,\chi'}(r,\eta)]_1 \frac{d\eta dr}{2\pi}.
\end{equation}
The integration over $d \chi'$ in  (\ref{2|1}) reflects the fact that
the intermediate state includes the coherent superposition
of the states of the pair with different $\chi'$
excited  due to inelastic collision.
The factor $ \exp [i \delta_1(\chi')]$  describes fast variation 
of the amplitude of the excitation with $\chi'$.

Integration in  (\ref{2|1}) over $d\eta$ is  
Gaussian, therefore it can be  fulfilled analytically, allowing 
the presentation of the  amplitude $S_{21}$ as a two-dimensional  
integral which can be easily handled numerically.

Knowing the matrix element $M$ one can present the cross-section 
of the reaction  (\ref{1}),(\ref{2}) 
in the form  
\begin{equation} \label{cs}
d\sigma_{\rm pci}(E) = \frac{ \Gamma_{\rm f} }{\Gamma}
R_{\rm pci} (E,\varepsilon)d\sigma_0(\varepsilon).
\end{equation}
The cross-section
$d\sigma_{\rm pci}(E)$ describes the full reaction in which PCI
plays a role.
The cross-section $ d\sigma_0(\varepsilon)$ describes 
only reaction  (\ref{1}) 
in which the role of PCI is neglected.
 The factor
\begin{equation} \label{REe}
R_{\rm pci}(E,\varepsilon) =  \frac{\Gamma}{2\pi}\left| S_{21}\right|^2
\end{equation}
presents the influence of PCI in the explicit form described by
the amplitude $S_{21} $  in  (\ref{2|1}).
This factor  has a resonant nature
sharply depending both on the above-threshold energy $\varepsilon$ and
the final-state energy of the pair $E$.

Equations (\ref{2|1}),(\ref{cs}),(\ref{REe})  
explicitly describe the influence of PCI on the cross-section.
The physical consequences of these formulae are discussed in the next
Section \ref{conseq}.

\section{Manifestations of PCI}
\label{conseq}
\subsection{Lorentz line}
Consider first the trivial situation when one neglects the effects of 
PCI
supposing that the Coulomb charge of the ion in the intermediate
and final state remains the same $Z_1=Z_2$. Then the wave functions
marked by symbols 1 and 2 in  (\ref{so}),(\ref{fswf}) differ only
by the momenta $k$ and $p$. Notice that the coefficients
$A(x), B(x),C(x)$ in the expansion of $\Sigma(r,\eta;p,\chi)$
 (\ref{abc}) are smooth functions. Therefore using inequality  (\ref{ge})
we can assume that they do not vary much
when the energy parameter $\epsilon$ in  (\ref{dx}) 
takes either the value
$\epsilon = E$, or $\epsilon = \varepsilon$, {\em i.e.}
$A(x)|_{\epsilon =E} \simeq A(x)|_{\epsilon =\varepsilon},~
B(x)|_{\epsilon =E} \simeq B(x)|_{\epsilon =\varepsilon} ,~
C(x)|_{\epsilon =E} \simeq C(x)|_{\epsilon=  \varepsilon}$.
The last equality greatly simplifies integration 
over $d\eta$ in  (\ref{2|1}) 
\[ \int \exp[ i B(x)(\chi-\chi')\eta]\frac{d \eta}{2\pi} = \frac{1}{B(x)}
\delta_D(\chi-\chi'). \]
Here $\delta_D(\chi-\chi')$ 
is the Dirac delta-function, which is eliminated by
the following integration over  $d\chi'$ in  (\ref{2|1}). As a result
the amplitude  can be presented as the simple one-dimensional integral
\begin{equation} \label{1d}
S_{21} =\exp [i\delta(\chi)] \int_0^\infty \phi^{(-)}_p(r; w(\chi))
\phi^{(+)}_k(r;w(\chi))dr
\end{equation}
with the Coulomb wave functions in the field of the
same effective Coulomb charge
$w(\chi)$. Evaluating it 
with the help of  (\ref{ca})
and substituting into  (\ref{REe}) one finds
that the resonant factor $R_{\rm pci} (E,\varepsilon)$ 
reduces to   the Lorentz line $ R_{\rm ll}(E,\varepsilon)$
\begin{equation} \label{Ll}
R_{\rm pci} (E,\varepsilon)\rightarrow R_{\rm ll}(E,\varepsilon) \equiv
\frac{\Gamma/2\pi}{ (E-\varepsilon)^2+ \Gamma^2/4},
\end{equation}
 as it should be when PCI  is neglected.
Notice that this result justifies the normalization of the wave functions
in  (\ref{ca}).
	
\subsection{Eikonal region}
\label{eikonal}

Let us examine now 
the influence  of PCI on the Wannier pair. 
 The integration over $dr$ in  (\ref{2|1})
is localized in the region $r\le r_{\rm l}$  (\ref{lo}).
It is essential how the localization radius $r_{\rm l}$
is related to the Coulomb radius $r_{\rm c}$  (\ref{er}).
Let us examine the situation when  condition 
\begin{equation} \label{>}
r_{\rm l} \equiv \frac{ 2 \sqrt{2 \varepsilon} }{\Gamma}\gg r_{\rm c}  
\equiv \frac{1}{E}
\end{equation} 
is fulfilled.
It is valid for sufficiently high energies $E,\varepsilon$.
In this case  the main contribution to the integral  (\ref{2|1})
comes from the region
where asymptotic relations
  (\ref{11}),(\ref{12}),(\ref{22}) are fulfilled.
They imply that $B_1(r,k)\simeq B_2(r,p) \simeq 1$ and
$C_1(r,k) \simeq C_2(r,p) \simeq 0$. These equalities
show that the integration over $d\eta d\chi'$ in  (\ref{2|1})  gives
the trivial result
\begin{equation}\label{tri}
\int \exp [i(\chi-\chi')\eta ]\frac{d\eta d\chi'}{2\pi} =1.
\end{equation}
Using it we find that the matrix element $S_{21}$ 
is simplified to be proportional to a  one-dimensional overlapping 
integral 
\begin{eqnarray}\label{rem}
S_{21} &=& \exp [i\delta_1(\chi)] \langle \phi_2 | \phi_1 \rangle,
\\ \label{pp}
\langle \phi_2 | \phi_1 \rangle &=&
  \int_0^\infty  
\phi^{(-)}_p(r; w_2(\chi))\phi^{(+)}_k(r;w_1(\chi)) dr .
\end{eqnarray}
Both wave functions 
here describe the  effective single-particle  Coulomb problem.
This problem  in the intermediate state is governed by the
effective charge $w_1(\chi)$ defined by  
(\ref{w}),(\ref{W02}),(\ref{w''})
with the ion  charge $Z=Z_1=1$.
In the final state the effective charge 
$w_2(\chi)$ depends on the final-state ion charge $Z=Z_2=2$.

From  (\ref{rem}),(\ref{REe}) we find the following simple
representation for the resonant factor
\begin{equation}\label{R0}
R_{\rm pci} (E,\varepsilon) \simeq R_{\rm c}(E,\varepsilon)
\equiv \frac{\Gamma}{2\pi} | \langle \phi_2 | \phi_1 \rangle |^2.
\end{equation} 
The   factor $R_{\rm c}(E,\varepsilon)$ defined in  (\ref{pp})
depends   on the overlapping integral between
the wave functions describing the effective
single-particle Coulomb problem.
It is remarkable that
this integral proves to be  
similar to the  quantity which was well-known previously 
in the traditional PCI problem  when a single low-energy 
electron  is influenced by the products of the autoionization decay.
In this case the shake-down model 
\cite{read2,shd,ks} 
gives correct description of PCI.
The shake-down model is based on an
overlapping integral with single-electron wave functions, whose
structure is very close  to the integral  in 
 (\ref{pp}),(\ref{R0}).

We come to  a very  interesting conclusion.  
Our knowledge of PCI properties of the traditional
situation when there is only one near-threshold electron
enables us to predict manifestations
of PCI when the Wannier pair is involved.
Conveniently we do not even need to 
fulfill explicitly calculation of the overlapping integral  (\ref{rem})
which appears in (\ref{R0}).
Inequality  (\ref{>}) guarantees that 
after the decay the Wannier pair is  located
so far away from the atom that neither the Coulomb charge
of the ion nor repulsion between the electrons
can significantly change the classical trajectory.
Thus the pair moves almost without acceleration, which
greatly simplify  behaviour of the wave functions
describing this motion.
A similar 
case was well studied in the single-electron problem. It was shown 
\cite{ks} that the eikonal approximation and 
the shake-down model give the same results when 
conditions of applicability for the eikonal theory are fulfilled.
Borrowing this result from the true single-electron problem 
and applying it to our effective single-particle problem
we {\em conclude }
that the eikonal theory
proves to be correct for the Wannier pair if condition
 (\ref{>}) is fulfilled. 
Notice that condition  (\ref{>}) can be rewritten 
in the standard in the eikonal 
approximation  form $ \varepsilon^{3/2}\gg \Gamma$.
Of course one could anticipate from the very beginning
that when this usual condition is valid, then the eikonal theory 
should give the correct description of PCI
for the Wannier pair. However, the peculiarity of the Wannier problem 
certainly appeals for a steady basis  for such  a  statement. 
That is why it is  important that 
we have been able to demonstrate validity of the eikonal theory by direct
calculations.

Recognizing that the eikonal approximation is valid, 
we conclude that PCI for the Wannier pair should have
the following  manifestations. \\
1. The line in the autoionization
spectrum is shifted by PCI. The shift is equal to
\cite{ks1,ks}
\begin{eqnarray}\label{eik}
\Delta E &=& - \frac{\Gamma}{2}\xi, \\ \label{xi}
\xi      &=& \frac{w_2-w_1}{\sqrt{2 \varepsilon}} = \frac{1}
{{\rm v}_\alpha}+ \frac{1}{{\rm v}_\beta}.
\end{eqnarray}
Here ${\rm v}_\alpha = {\rm v}_\beta = \sqrt{\varepsilon}$. 
The right-hand side in  (\ref{xi})
is presented in the form which is common in the eikonal approximation.\\
2. The line changes the shape, becoming broader, lower, and 
acquiring  the   ``left shoulder'' which makes it asymmetrical. 
All these variations are the more significant the larger is the 
basic parameter
$\xi$. These properties of the line are described by the profile
 \cite{ks1,ks} 
\begin{eqnarray}\label{erf}
&&R_{\rm pci}(E,\varepsilon) \simeq R_{\rm eik}(E,\varepsilon)= 
\frac{\Gamma/2\pi}{ (E-\varepsilon)^2+\Gamma^2/4}
k_{\rm eik}( (E-\varepsilon)/\Gamma,\xi),
\\ \nonumber
&&k_{\rm eik}( (E-\varepsilon)/\Gamma,\xi) \equiv
\frac{\pi \xi}{\sinh \pi\xi}
\exp \left( -2 \xi \arctan \frac {E-\varepsilon}{\Gamma/2} \right).
\end{eqnarray}
3. In spite  of strong variations in the shape  and position the
intensity of the line remains constant,  it is not
influenced by PCI, 
\begin{equation} \label{GG}
\int R_{\rm pci}(E,\varepsilon) dE =1.
\end{equation}
This property is essential
for the Wannier problem where intensity is one of the main
characteristics of the process.
Taking into account that the cross-section of the reaction
(\ref{1}) satisfies the power law (\ref{w1}) with the index
$\mu = \mu_1 = 1.127$ we find 
from (\ref{w1})(\ref{cs}),(\ref{GG}) that this law remains valid
for the total intensity of the line 
for reaction (\ref{1}),(\ref{2})
\begin{equation}\label{twp}
\sigma = 
\int \frac{d\sigma_{\rm pci}(E)}{d E} dE \sim \sigma_0
\sim \varepsilon^{\mu_1}
\end{equation}
considered as a function of the above-threshold energy.

Up to this point our consideration was restricted by two
particular assumptions. Firstly,
we considered the
case of large velocities of the autoionized electron  (\ref{3>})
which allows one to assume that it does not
play a role in PCI.
Secondly, we discussed  the opposite directions for the Wannier electrons
neglecting fluctuations in their angular distribution.
Once we have established that the eikonal theory is correct
we can remove both these simplifications and
include into our scheme the case of sufficiently low-energy
autoionization and arbitrary directions for the Wannier electrons.
As  usual in the eikonal approximation it is 
sufficient to redefine 
the parameter $\xi$ which has  the following more general form 
\begin{equation} \label{axi}
\xi = \frac{1}{{\rm v}_\alpha}+\frac{1}{{\rm v}_\beta}-
\frac{1}{{\rm v}_{\alpha\gamma}}
-\frac{1}{{\rm v}_{\beta\gamma}}.
\end{equation}
Here ${\bf v}_{\alpha\gamma} = {\bf v}_\alpha-{\bf v}_\gamma,~
{\bf v}_{\beta\gamma} = {\bf v}_\beta-{\bf v}_\gamma$.
With this parameter $\xi$
 (\ref{eik}),(\ref{erf}) describe the line which depends
on the absolute values as well as directions of 
the velocities of all three electrons.

\subsection{Coulomb region}
\label{coul}
Consider the case when the localization radius
$r_{\rm l}$  (\ref{lo}) is smaller than the Coulomb radius
$r_{\rm c}$  (\ref{er})
\begin{equation} \label{cc}
r_{\rm l} \equiv \frac{ 2 \sqrt{2 \varepsilon} }{\Gamma}
\ll r_{\rm c}  \equiv \frac{1}{E}.
\end{equation}
This inequality holds for sufficiently low energies $E,\varepsilon$.
This inequality means that the decay    takes place when the pair is 
so close to the ion that the Coulomb potential has a strong influence on 
trajectories of the Wannier pair.

Let us examine the behaviour of the integrand in the amplitude
$S_{21}$ in  (\ref{2|1})
as a function of $\chi'$. 
Remember that in the region $r \ll r_{\rm c}$ the asymptotic relations
 (\ref{aScc}),(\ref{aSpc}), and (\ref{aSp}) 
for the coefficients governing the action $\Sigma$ are valid.
Notice first of all that the quadratic in $\chi'$
terms in the exponent arising from the phase shift 
$\Delta \delta_1(\chi')$ and
from the term $[\Sigma(r,\eta;p,\chi')]_1$ cancel each other due to
 (\ref{aScc}),(\ref{Dd}). 
Another simplification comes from the fact that
for  small distances  the Coulomb function smoothly depends on 
$\chi'$ exhibiting    behaviour $ \sim \chi'^2$,
see discussion after  (\ref{sp}). Therefore 
one can suppose that
\[ 
[ \exp [i \delta_{\rm c} (\chi')]\phi^{(+)}_k(r;w(\chi'))]_1 \simeq
[ \exp [i \delta_{\rm c} (\chi)]\phi^{(+)}_k(r;w(\chi)) ]_1.
\]
These properties of the integrand 
allow analytical integration over $d\chi'$ 
in  (\ref{2|1})
\begin{equation}\label{chi'}
\int \exp (-i [ B(x)]_1 \chi' \eta) \frac{d\chi'}{2 \pi } = 
\frac{1}{ [ B(x) ]_1 } \delta_D(\eta).
\end{equation}
The delta-function is eliminated by the subsequent integration over 
$d\eta$.
As a result one finds the following simple representation for
the amplitude
\begin{equation}\label{S21}
\label{cS}
  S_{21}  =  \exp [ i \delta_{c,1}(\chi) ]
\langle \phi_2  | ( B_2/B_1 )^{1/2} | \phi_1 \rangle .
\end{equation}
Here the matrix element is defined as
\begin{equation}\label{meBB}
\fl \langle \phi_2 | ( B_2/B_1 )^{1/2} | \phi_1 \rangle = 
  \int_0^\infty  \phi^{(-)}_p(r;W_2(\chi) )
      ([B(x)]_2 / [B(x)]_1)^{1/2} 
      \phi^{(-)}_k(r;W_1(\chi)) dr. 
\end{equation}
Using  (\ref{cS}) we find the following
representation for the resonant factor  (\ref{REe})
\begin{equation}\label{Rpcic}
R_{\rm pci}(E,\varepsilon) = \frac{\Gamma}{2\pi}
\left| \langle \phi_2 |
\sqrt{B_2 / B_1}  | \phi _1 \rangle \right| ^2.
\end{equation}
This expression can be simplified even further.
The functions $[ B(x)]_1$  and $[B(x)]_2$ 
smoothly depend on the coordinate $r$.
In contrast, the Coulomb wave functions $|\phi_1\rangle, ~|\phi_2\rangle$
oscillate. Calculating the integral with
sufficiently fast oscillating functions one can use
the saddle-point method. The semiclassical phases of the wave 
functions
$|\phi_1\rangle, ~|\phi_2\rangle$ are 
$\int \sqrt{2( \varepsilon+i\Gamma/2+W_1(\chi)/r)}dr$ and 
$\int \sqrt{2(E +W_2(\chi)/r)}dr$ respectively. Their
difference is
\begin{equation}\label{phase}
\Phi(r) = \int^r \left[ \sqrt{2( \varepsilon+i\Gamma/2+
W_1(\chi)/r)} -  \sqrt{2(E +W_2(\chi)/r) } \right] dr.
\end{equation}
The saddle point
$r = r_{\rm sp}$ is to satisfy equation $\Phi'(r_{\rm sp}) 
=0$
yielding 
\begin{equation}\label{stat}
E+\frac{W_2(\chi)}{r_{\rm sp}} = \varepsilon + i\frac{\Gamma}{2}+
\frac{ W_1(\chi) }{r_{\rm sp}}.
\end{equation}
Solving this equation one finds
\begin{equation}\label{cro}
r_{\rm sp} = \frac{W_2(\chi) - W_1(\chi)}{\varepsilon - E+i\Gamma/2}.
\end{equation}
The energies $E,\varepsilon$ are low (\ref{cc}),
therefore the radius  $r = r_{\rm sp}$ is large. This 
fact justifies both the applicability of the semiclassical estimation
for the phases in  (\ref{phase}) and the validity of the 
saddle-point approximation. 
These arguments  demonstrate that calculating the matrix element
 (\ref{meBB}) we can suppose that
the main contribution to the
integral comes
from the vicinity of the saddle point  (\ref{cro}). This  allows one to
evaluate smooth functions $[B(x)]_1,[B(x)]_2$ at this point and
take them out of integration. As a result  
the resonant factor given by   (\ref{Rpcic}) can be simplified 
to the following final form which was first announced in (\ref{rkr})
as the  major result of this paper
\begin{equation}\label{pcif}
R_{\rm pci}(E,\varepsilon) = K_{\rm w}(E,\varepsilon)
R_{\rm c}(E,\varepsilon),
\end{equation}
where the  factor 
\begin{equation}\label{ov2}
R_{\rm c}(E,\varepsilon) \equiv \frac{\Gamma}{2 \pi}
\left| \langle \phi_2 | \phi _1 \rangle \right| ^2
\end{equation}
depends  on the overlapping integral 
$ \langle \phi_2 | \phi _1 \rangle $. 
Notice that this  integral can   be presented in an analytical form
if the integration is pushed one step further
using the saddle-point method, but for our
purposes it is sufficient to  keep  it in general form.

We met the factor  (\ref{ov2}) before 
when discussed
the eikonal approximation, see  (\ref{R0}).
It was argued there that
the factor $R_{\rm c}(E,\varepsilon)$
describes those properties of PCI   which
were previously well known from traditional  studies of  
PCI with one low-energy electron. 
In contrast, the factor
\begin{equation}\label{kw}
K_{\rm w}(E,\varepsilon) = \left| \frac{B_2(x_{2,sp})}{B_1(x_{1,sp})} 
\right|
\end{equation}
which appears in  (\ref{pcif}) is a  new quantity
specific for PCI with Wannier electrons, having no
analogue in  PCI with
one low-energy electron.
Due to this reason it is called the Wannier factor.
Remember that the function $B(x)$  defined by 
(\ref{S2si2}),(\ref{solchi}),(\ref{dx})
gives normalization for 
the wave function of  Wannier
electrons   (\ref{answ}). 
The found 
Wannier factor depends on a ratio of 
these normalization functions for
the intermediate and final states.
The  coordinates of normalization functions, 
in accordance with  (\ref{dx})
are defined   as
\begin{eqnarray}\label{x1}
x_{1,sp} &=& \frac{1}{2}\left[ \left( 1+ \frac{W_1(\chi)}{\varepsilon 
r_{\rm sp}} 
\right)^{1/2} +1 \right], \\ \label{x2} 
x_{2,sp} &=& \frac{1}{2}\left[ \left( 1+ \frac{W_2(\chi)}{E r_{\rm sp}}
\right)^{1/2} +1 \right].
\end{eqnarray}
Traditionally the interest in the Wannier problem
has been inspired by the near-threshold power-law of the
cross-section  (\ref{w1}). 
It originates from the  asymptotic behaviour
of the normalization coefficient $B(x)$
for small separations when $E r \ll 1$, see  (\ref{aSpc}). 
Thus traditional studies of the Wannier problem can be considered
as a probing of the
normalization coefficient $B(x)$ in the particular
asymptotic region well inside the Coulomb zone.
PCI opens   an interesting opportunity 
to investigate this coefficient in  a much  wider area,
not necessarily restricted by the Coulomb zone.
According to  (\ref{x1}),(\ref{x2})
variation of the above-threshold energy $\varepsilon$ and the final-state
energy $E$ leads to variation of  $x_{1,sp},~x_{2,sp}$ 
in a broad area resulting in significant variation of normalization
coefficients $[B(x)]_1$ and $[B(x)]_2$.

The result obtained  (\ref{pcif}) permits one to address
the question of the threshold behaviour 
of the cross-section
which is of particular interest for the Wannier problem.
When the energy of the Wannier pair is low $E\rightarrow 0$ 
then one finds from  (\ref{cro}),(\ref{x2})
that $r_{\rm sp} \simeq (W_2(\chi)-W_1(\chi))/\varepsilon$ and
$x_{2,sp} \sim 1/\sqrt E \rightarrow \infty$. The later
condition  allows the  use of asymptotic relation
 (\ref{aSpc}) which shows that $B_2(x_{2,sp}) \sim E^{\mu_2},~~
\mu_2 = \nu_2/2$.
From  (\ref{kw}) one finds the  same estimate  for the 
Wannier factor
\begin{equation}\label{asw}
R_{\rm pci}(E,\varepsilon) \sim E^{\mu_2}. 
\end{equation}
Equation  (\ref{pcif}) gives the same estimate for the resonant
factor 
$ R_{\rm pci}(E,\varepsilon) \sim E^{\mu_2}$ resulting in 
the threshold law for the differential cross-section as a function of the
energy of the pair $E$
\begin{equation}\label{mu2}
\left( \frac{ d \sigma}{d E}\right)_{pci} \sim E^{\mu_2},~~~~E\rightarrow
 0.
\end{equation}
Similarly one can estimate the behaviour of the resonant factor
on the above-threshold energy 
$ R_{\rm pci}(E,\varepsilon) \sim \varepsilon^{-\mu_1}, ~~\mu_1=
\nu_1/2$. 
Remembering that the cross-section $\sigma_0$ of the 
exhibits the usual Wannier power-type behaviour 
$\sigma_0 \sim  \varepsilon^{\mu_1},~~~~\varepsilon \rightarrow 0$
as a function of the above-threshold  energy
we find that the cross-section for the combined process 
 (\ref{1}),(\ref{2}) does {\em  not} depend on the 
above-threshold energy $\varepsilon$
\begin{equation}\label{con}
\left( \frac{ d\sigma}{dE}\right)_{pci} \sim const,~~~~\varepsilon
\rightarrow 0.
\end{equation}
We see that PCI has a dramatic effect on the threshold behaviour
of the cross-section.
Firstly, the PCI results in the variation of the exponent.
The exponent in  (\ref{mu2}) corresponds to the final-state 
charge of the ion $Z=Z_2=2$ which gives $\mu_2 = 1.056$.
If the Auger decay is impossible,
then the cross-section of the reaction  (\ref{1})
is described by the usual Wannier power law
 (\ref{w1}) which corresponds to 
the charge of the ion $Z=Z_1 =1$ resulting in $\mu_1 = 1.127$.
Secondly, the cross-section does not  depend on the above-threshold
energy $\varepsilon$. 
These results agree  qualitatively with the  simple physical picture.
If above threshold energy is low then 
the decay takes place when the pair
is so close to the ion that its separation 
 can be neglected  compared with the large radius of the Coulomb
zone.  This means that
the propagation of the pair in the
intermediate state is insignificant, which 
makes insignificant as well the very existence of this intermediate 
state.
That is why 
the final answer for the Wannier exponent should not depend
on parameters governing the intermediate state.
The fact that  (\ref{pcif}) reproduces
this result can be considered as  a qualitative 
verification of this formula.

Generally speaking,
one could contemplate a possibility to
 measure the power-type behaviour   (\ref{mu2}).
To this end one should
fix the  above-threshold energy and 
measure  the ``left shoulder'' of the resonance
profile in the near-threshold region versus the energy of the Wannier 
pair.
However this project meets  a difficulty. 
It can be verified that in order  to distinguish the exponent 
$\mu_2$ from $\mu_1$ in the power-type behaviour  of the cross-section
 (\ref{mu2}) the left-hand side in  inequality (\ref{cc})
should be really small, say, less than $\le 10^{-2}$ putting
a  severe restriction on $E,\varepsilon$. 

The most important result of this Subsection is 
 (\ref{pcif}) which  describes the influence of PCI on the resonance 
profile.
This equation  was evaluated assuming that energies $E,\varepsilon$ are
low enough to satisfy condition  (\ref{cc}). Notice, however, that
the equation  remains correct for much higher energies satisfying
the condition of applicability of the eikonal approximation
 (\ref{>}) which is opposite to the low-energy limit.
To see this consider the eikonal region  (\ref{>}) in which
$E \sim \varepsilon\sim |E-\varepsilon| \gg
\Gamma^{2/3}$. For these energies   (\ref{cro})
results in $r_{\rm sp}\sim \Gamma^{-2/3}$ and therefore 
$E r_{\rm sp} \sim \varepsilon r_{\rm sp} \gg 1$.
Equations 
 (\ref{x1}),(\ref{x2}) show
that $x_{1,sp}\simeq x_{2,sp} \simeq 1$
permitting one 
to use asymptotic condition  (\ref{12})
which shows  that the Wannier factor
is  trivial $K_{\rm w}(E,\varepsilon) \simeq 1$. Therefore,
in the eikonal region  (\ref{pcif}) yields
$R_{\rm pci}(E,\varepsilon) \simeq R_{\rm c}(E,\varepsilon) $
which according to  (\ref{R0}) is the correct result.
The fact that  (\ref{pcif}) proves to be correct
in both the eikonal region $r_{\rm l} \ll  r_{\rm c}$ and
the Coulomb region $r_{\rm l} \gg r_{\rm c}$ may be considered as 
an indication that it should give reasonable results
in the intermediate region $r_{\rm l} \sim r_{\rm c}$ as well.

\section{Electron impact in the vicinity of Ne K-shell}
\label{examp}
In order to illustrate the validity of  different approximations
developed above for description of PCI
let us consider the example studied experimentally 
in \cite{mel} in which  
the electron impact ionization
of the Ne atom in the vicinity of the K-shell
\begin{equation}
{\rm e} + {\rm Ne} \rightarrow {\rm e+e+Ne}^+(1s^{-1})
\end{equation}
is followed by the KLL Auger  decay 
\begin{equation}
{\rm Ne}^{+}(1s^{-1}) \rightarrow 
{\rm e} + {\rm Ne}^{2+}(2p^{-2}).
\end{equation}
 
The distortion of resonance profiles  by PCI
is illustrated in figure 1 for 
above-threshold energies
$\varepsilon =2,3,5$ and $10$ eV.  
The width of the K-vacancy is 
$\Gamma = 270$ meV \cite{kraus}.
The profiles are calculated for 
the symmetric geometry of   Wannier electrons when they move
in opposite directions  possessing equal  energies.
Notice that experimentally \cite{mel} 
the spectra were measured versus the energy of the Auger electron
when neither energy nor angular distributions of Wannier electrons were 
fixed.

The full curve in figure 1 presents the results of calculations 
based on  
the most advanced equation (\ref{pcif}) which, as was argued above 
is valid for both near-threshold  and high energy regions.
The factor $R_{\rm c}(E,\varepsilon) = \Gamma/(2\pi)
|\langle \phi_2 | \phi_1 \rangle|^2 $ is  calculated 
by direct numerical integration of the overlapping integral with
the Coulomb wave functions describing
the effective single-particle problem. 
Remember that this problem
gives the simplest description of the Wannier pair.
Normalization of the Coulomb wave functions
is given in  (\ref{ca}). 
For numerical calculations
it is more convenient to work with the real Coulomb wave function
$\phi_2(r)$ taking a linear combination of convergent and divergent
waves,  rather then only the convergent wave.
This is possible because
the divergent wave in $\phi_2(r)$ does not contribute
to the resonance, see discussion before (\ref{fswf}).
The wave function $\phi_1(r)$ has to be taken as a divergent
wave localized in the vicinity of the atom \cite{ks}.
The Wannier factor $K_{\rm w}(E,\varepsilon)$
is found from (\ref{S2si2}),(\ref{solchi}).

Full dotes in figure 1 present calculation based on (\ref{Rpcic}).
The perfect agreement between the full lines  and the  full circles  shows
the high accuracy of the  approximation which was used in order to present
the resonant factor   in the final form  (\ref{pcif})
as a product of two different factors.

Dotted lines in figure 1 present the simple eikonal
approximation  (\ref{erf}) which should be correct only  for
high energies. 
Results presented 
show that below $\varepsilon = 5$ eV the eikonal approximation fails,
while for high above-threshold energies $\varepsilon \ge 10$ eV 
it produces quite reliable results.  Notice that 
experimental results of \cite{mel} belong to the region 
$\varepsilon \ge 10$ eV
where we have just verified  the eikonal approximation to be  valid.

The dotted line in figure 1({\it a}) shows the
Coulomb factor $R_{\rm c}(E,\varepsilon)$ which is known to produce
good results for PCI with a single  low-energy electron.
Figure (1{\it a}) shows 
that for the considered situation with two low-energy electrons
it proves to  be incorrect for low energies. This  demonstrates
that the Wannier factor $K_{\rm w}(E,\varepsilon)$ plays a very
important role.
The Wannier factor  exhibits quite an interesting 
variation as  demonstrated in figure 2 where  it
is presented versus the energy of the 
pair $E$ for the fixed above-threshold energy $\varepsilon = 2$ eV.
In order to reveal the asymptotic low-energy behaviour  
$\sim E^{\mu_2}, ~~E\rightarrow 0$ predicted by  (\ref{mu2})
the Wannier factor is shown as $K_{\rm w}(E,\varepsilon)/E$.

The PCI shift found from calculations of the PCI profiles 
is shown in figure 3 for the symmetrical configuration
of Wannier electrons.
The full curve gives the results obtained from (\ref{pcif}).
The dotted line shows the prediction of the eikonal
approximation.
The full dottes in  figure 3  present the results extracted from the 
profiles  calculated in  \cite{ost}.
These results effectively include different possible
configurations of Wannier electrons in  the final state.
One can expect that the symmetrical configuration 
of the Wannier
electrons gives large contribution to the total
probability. Therefore the results 
calculated for the symmetrical  configuration 
should reasonably agree with calculations in which all
continuum states of Wannier electrons  are included.
Notice that the theory developed above and the approach 
of \cite{ost} are very different in technique used.
Nevertheless there is a clear resemblance
in several important basic features. 
Firstly, in both works PCI is described with the help
of overlapping integrals between wave functions describing the
Wannier pair. Secondly, both works 
rely upon the semiclassical approximation.
Having in mind these basic similarities
one could expect that  
the two approaches should give similar
results. This hope proves be correct for high energies, 
see figure 1({\it d}) for 10 eV above the threshold.
There is, however, some discrepancy below 10 eV. 
It can be verified comparing figure 1
with the  results presented in figure of \cite{ost}
that  distinctions exist  in profile shapes as well.

\section{Discussion of the main results}
\label{discussion}
This paper develops the following ideas and results.

1.It is emphasized  that the 
propagation of two Wannier electrons can be described by an effective
single-particle Coulomb problem
with corrections which take into account 
more subtle three-body properties.
Correspondingly 
the wave function of the Wannier pair (\ref{answ}) is presented as 
the  product of the Coulomb wave function describing  the effective
single-particle Coulomb problem, 
the additional phase factor $\exp(i\Sigma)$, and
the normalization coefficient $B$ (\ref{S2si2})  
which plays a very important
role in our consideration.

2.The complex  problem 
(\ref{1}),(\ref{2})
with three electrons in the final state is reduced to a more
simple problem in which only  two Wannier electrons play a role.
The key point is an assumption that the autoionization
energy is so high that the autoionized electron cannot 
interact strongly with the other two electrons.
The dynamical properties responsible for PCI 
are shown to be described by the overlapping integral
(\ref{2|1}) between the wave functions of
Wannier electrons in the final and intermediate
states.

3. The resonant factor is found in a  very simple form
(\ref{pcif}) which depends on two factors. One
of them $R_{\rm c}$ describes the overlapping integral 
between the Coulomb wave functions for the effective
single-particle problem. A similar integral is known 
to describe PCI when there is only one near-threshold electron.
More specific properties of the considered problem are
described by the Wannier factor $K_{\rm w}$.
The factorization of the resonant factor into a product
of $K_{\rm w}$ and $R_{\rm c}$ has a clear physical reason.
There are two scales in the problem.
The effective single-particle Coulomb problem describes strong
variations of the wave function with  a typical 
radius $r_{\rm single}\sim 1/p$. In contrast, the quantities 
$\Sigma, B$ describing
the specific properties of the three-body problem vary
with a typical distance  equal to the radius of the Coulomb
zone $r_{\rm 3-body} \sim 1/p^2$. For low energies
these radii are different $r_{\rm single} \ll r_{\rm 3-body}$.
It is well known that when there are two different scales
in a problem  then
one has to expect an amplitude and a probability to
be presented as products of quantities describing 
events which happen  in different scales, as it happens in our case.

Equation (\ref{pcif}) was verified for low energies, {\it i.e.}
in the Coulomb region, and for high energies in the eikonal zone.
For intermediate region one can {\it  hope} that it should 
give reasonable results. 
(However, validity of the later 
assumption  should be clarified by numerical 
integration of the matrix element (\ref{2|1}).)

The Wannier factor $K_{\rm w}$ is shown to depend on the normalization 
coefficient $B$ of the wave function of the Wannier pair, 
see (\ref{kw}). The distance at which this coefficient
is to be taken depends on the above-threshold energy as well
as on the energy of the pair in the final state, see
(\ref{cro}). It is remarkable that the same coefficient
considered for small separations (\ref{aSpc})
governs the Wannier power law (\ref{w1}),
and therefore it has been under 
thorough experimental
investigation for a long period of time.
Equation (\ref{pcif}) shows that there is  a new possibility
to experimentally investigate the  coefficient $B$ in
a wide range of distances.

4.It is demonstrated that for low energies  of the pair $E\rightarrow 0$
the Wannier power law has the form (\ref{asw}) with 
the exponent governed by the ion charge in the final state.
In contrast for higher energies
in the eikonal region the power law has a form of (\ref{twp})
depending on the above-threshold energy $\varepsilon$ with the exponent 
governed by the ion charge in the intermediate state.

5.Our consideration was restricted by 
the one-dimensional model formulated in Section \ref{wan}.
However, one can hope that the approach developed  is more general 
than the model itself.
There are   clear  physical reasons for this. 
Firstly, PCI as well as the Wannier
problem manifest themselves
for large separations  where the Coulomb
potential play an important  role, while the more rapidly decreasing
orbital barrier does not. Therefore  neglect
of this later potential seems reasonable.
Secondly, the arguments considered above show that the 
factorization of the resonant factor
into the product of the Coulomb factor and the Wannier factor is
a general property caused by existence of the two different
scales in the problem. Therefore it should remain valid
when the angular variables are included as well.

\ack
I am thankful to V.N.Ostrovsky who drawn my attention to the problem
discussed in this paper providing  me with the preprint of 
the paper \cite{ost}. 
The support of the Australian Research Council is acknowledged.


\Figures

\begin{figure}
\caption{Electron impact with excitation of
K-shell in Ne.  Calculated
profiles of Auger lines distorted by PCI $\protect R(E,\varepsilon)$
versus the energy of the Wannier pair $\protect E$ for a fixed 
above-threshold
energy $\protect \varepsilon$. 
The width of the K-vacancy is 270 meV \protect \cite{kraus}. 
Figures ({\it a}),({\it b}),({\it c}),and ({\it d}) 
present results for $\protect \varepsilon=2,3,5$
and $\protect 10$ eV.
The symmetric configuration of Wannier
electrons is considered, they move in opposite directions possessing
equal energies.
\chain~ the non-distorted Lorentz line,
\full~ prediction  of the most advanced equation (\protect \ref{pcif}),
\longbroken~ the eikonal approximation 
$\protect R_{\rm eik}(E,\varepsilon)$
(\protect \ref{erf}),
\protect \fullcirc~  prediction of equation (\protect \ref{Rpcic}),
\dashed~  the Coulomb factor $\protect R_{\rm c}(E,\varepsilon)$
(\protect \ref{ov2}).
The perfect agreement between the full line and full-circled one shows
the high accuracy of the  approximation which permits 
to present the resonant factor in the final form  
(\protect \ref{pcif}).
Figure ({\it d}) shows that 
the eikonal approximation is valid above 10 eV. The strong deviation 
of the dashed line from the full one in figure ({\it a}) indicates an
important role played by the Wannier factor $\protect 
K_{\rm w}(E,\varepsilon)$
for low energies.}
\end{figure}

\Figure{
The same reaction as in figure 1.
\full 
the ratio of the Wannier factor  $\protect
K_{\rm w}(E,\varepsilon)$ to the energy
of the pair $\protect E$ 
versus the energy $\protect E$  for the fixed above-threshold
energy $\protect \varepsilon =2$ eV,
\broken  low-energy asymptotic behaviour (\protect \ref{mu2}).}

\Figure{The same reaction as in figure 1.
The calculated PCI shift versus the above-threshold energy:
\full  prediction based on equation (\protect \ref{pcif}),
\broken  the eikonal approximation (\protect \ref{erf}),
\protect \fullcirc results extracted from profiles calculated in 
\protect\cite{ost}.}

\begin{figure}[t]
\input psfig
\psfig{file=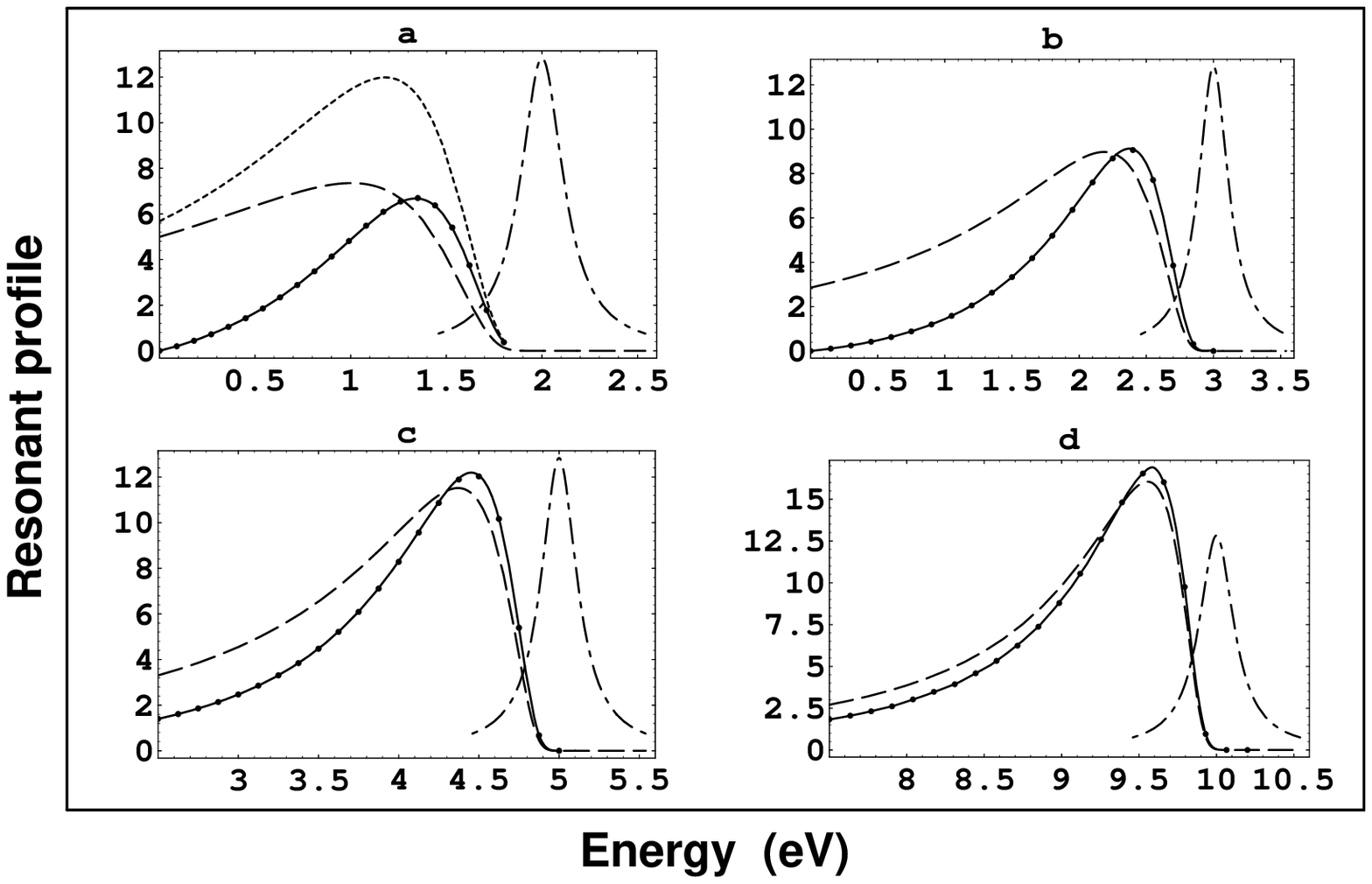}
\hspace{68mm}
{\huge Figure 1}
\end{figure}

\begin{figure}[t]
\input psfig
\psfig{file=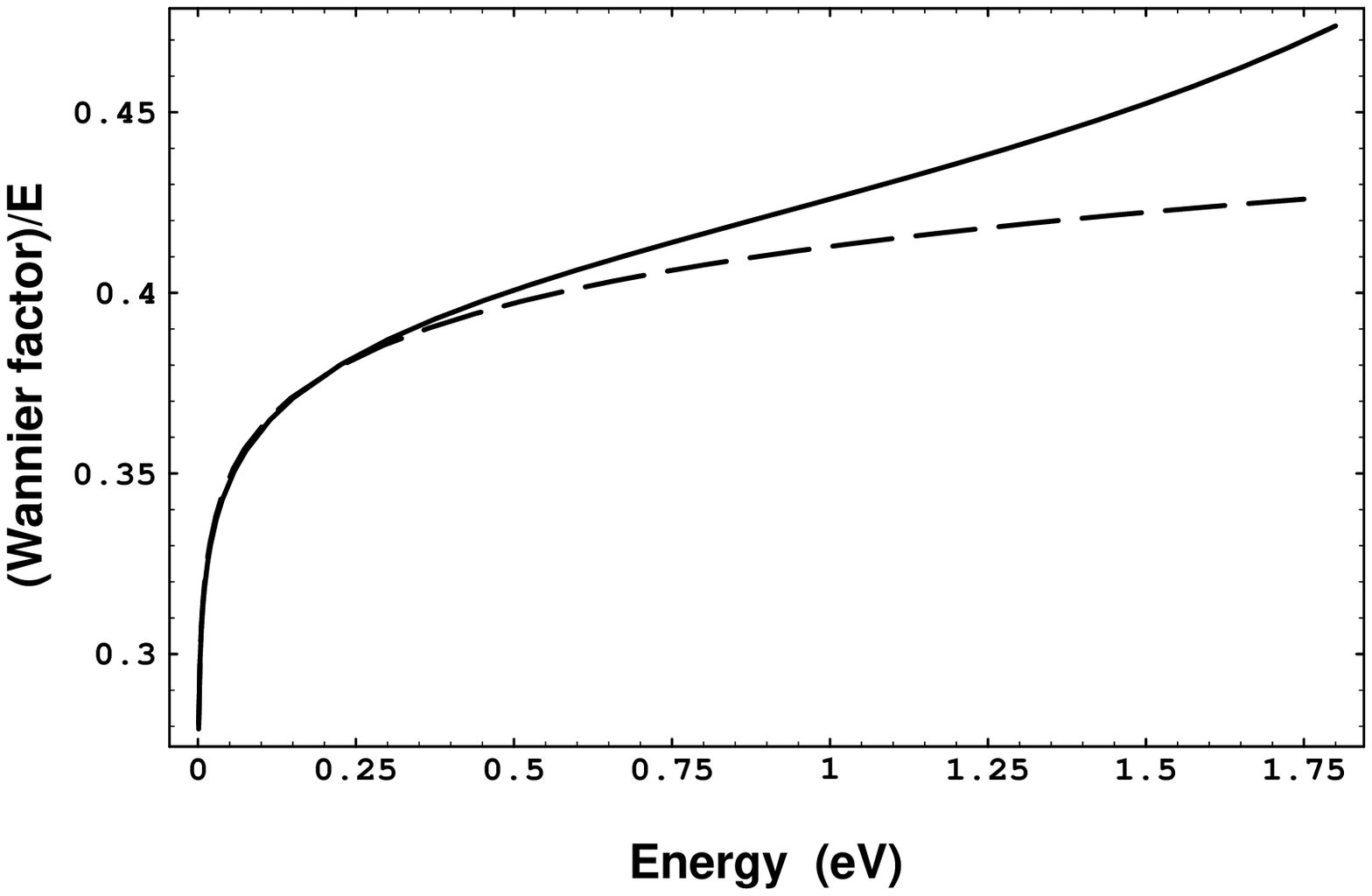, clip=}
\hspace{75mm}
{\huge Figure 2}
\end{figure}

\begin{figure}[t]
\input psfig
\psfig{file=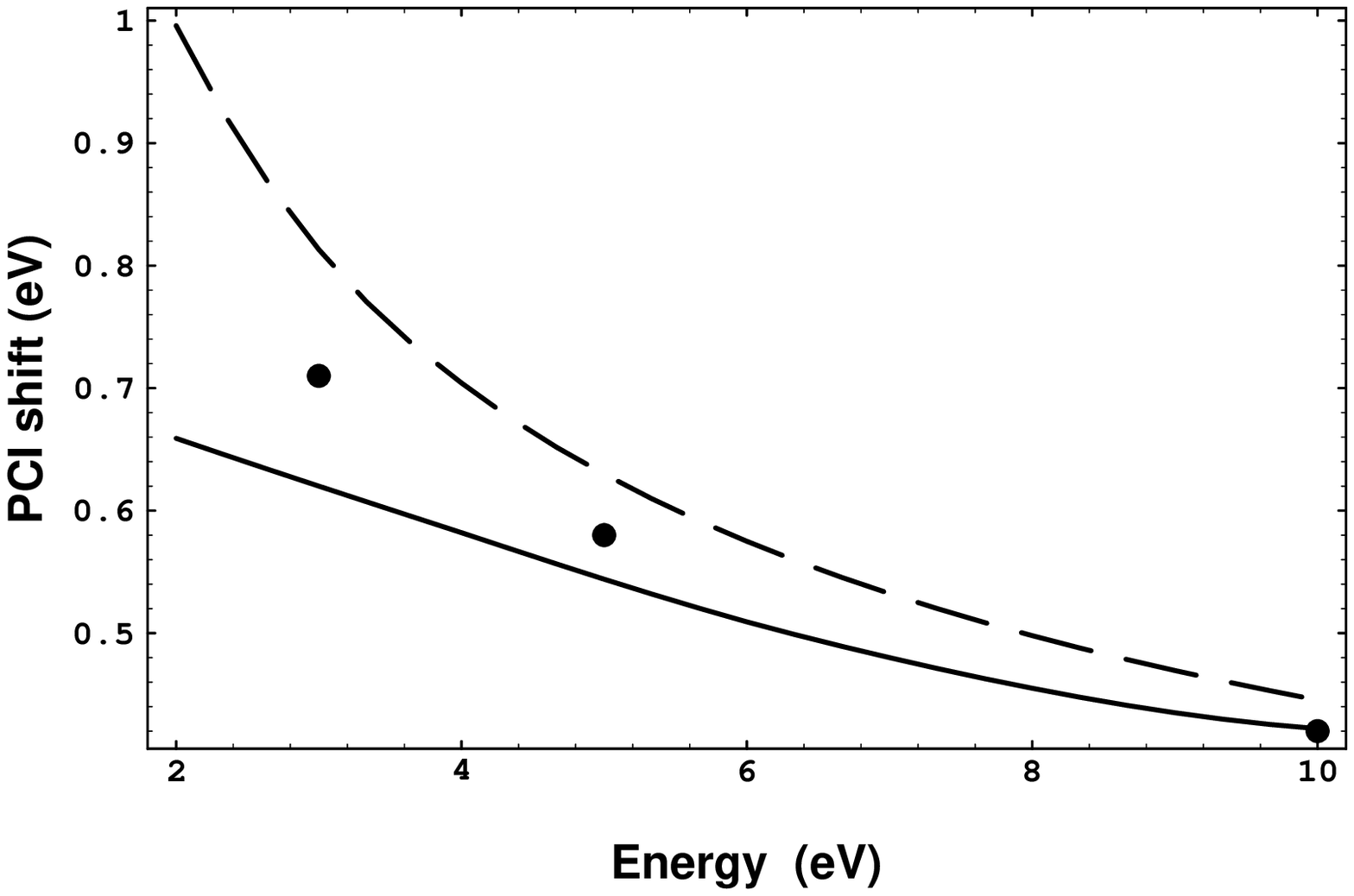, clip=}
\hspace{75mm}
{\huge Figure 3}
\end{figure}


\Bibliography{18}


\bibitem{mel} Kamm M, Weber W and Mehlhorn W 1994
{\it J.Phys.B:At.Mol.Opt.Phys.}  {\bf 27} 2585 

\bibitem{ost} Kazansky A K and Ostrovsky V N 1996
{\it 5-th International Workshop ``Autoionization Phenomena in Atoms'',
Eds.V.V.Balashov, A.A. Grum-Grzhimailo, and E.A.Romanovsky (Moscow)}
 (Moscow University Press)  p~67


\bibitem{wan} Wannier G H 1953 {\it Phys.Rev.} {\bf 90} 817 

\bibitem{pet} Peterkop R 1971   {\it J.Phys.B:At.Mol.Opt.Phys.}  
{\bf 4} 513 

\bibitem{rau} Rau A R P 1971 {\it Phys.Rev.A} {\bf 4} 207 

\bibitem{kla} Klar H and Schlecht W  1976 {\it J.Phys.B:At.Mol.Opt.Phys.}
{\bf 9} 1699 


\bibitem{wat} Watanabe S 1987 {\it Phys.Rev.}A {\bf 36} 1566

\bibitem{ost2} Kazansky A K and Ostrovsky V N 1992
{\it J.Phys.B:At.Mol.Opt.Phys} {\bf 25} 2121 

\bibitem{ks} Kuchiev M Yu and Sheinerman S A  1989   
{\it Sov.Phys.Uspechi} {\bf 32} 569

\bibitem{petlie}
Peterkop R and Liepinsh 1969
{\it Abstr. 6th Int. Conf on the Physics of Electronic and Atomic 
Collisions} (Cambridge, Mass: M.I.T. Press) 212

\bibitem{lan} Landau L D  and Lifshits  E M  1977 
{\it Quantum mechanics     
3d ed., rev. and enl. Course of theoretical physics; v.3.}
( Oxford ; New York : Pergamon Press)

\bibitem{fo}  Fock V A 1978
{Fundamentals of quantum mechanics}  (Moscow : Mir) 


\bibitem{read2}
Read F 1975 {\it Rad.Res.} {\bf 64} 23

\bibitem{shd} King G C , Read F H and Bradford R C
1975  {\it J.Phys.B:At.Mol.Opt.Phys} {\bf 8} 2210 

\bibitem{ks1}Kuchiev M Yu and  Sheinerman S A  1986
 {\it Sov.Phys.-JETP}  {\bf 63} 986 

\bibitem{kraus}
Krause M O and Oliver J H 1979 {\it J.Chem.Phys.Ref.Data} {\bf 8} 329 
\endbib
\end{document}